\pdfoutput=1
\documentclass[12pt,a4paper]{article}
\usepackage{ifthen} % for conditional statements
\newboolean{pdflatex}
\setboolean{pdflatex}{true} % False for eps figures 

\newboolean{articletitles}
\setboolean{articletitles}{true} % False removes titles in references

\newboolean{uprightparticles}
\setboolean{uprightparticles}{true} %True for upright particle symbols

\def\paperauthors{LHCb collaboration} % Leave as is for PAPER, CONF and FIGURE
\def\paperasciititle{Search for heavy neutral leptons in B-meson decays} % Set ASCII title here !! MAKE sure it's only ASCII characters !! 
\def\papertitle{Search for heavy neutral leptons in B-meson decays} % Latex formatted title
\def\paperkeywords{{high energy physics}, {LHCb}, {heavy neutral leptons}, {HNL}, {FIP}, {long-lived particle}, {LLP}, {HNLs}, {FIPs}, {long-lived particles}, {LLPs}} % Comma separated list
\def\papercopyright{\the\year\ CERN for the benefit of the LHCb collaboration} % new since 9/Apr/2018
\def\paperlicence{CC BY 4.0 licence}
\def\paperlicenceurl{https://creativecommons.org/licenses/by/4.0/}

\newif\ifEnableSectionTOCLinks
\EnableSectionTOCLinksfalse % deactivated

\usepackage[top=1in, bottom=1.25in, left=1in, right=1in]{geometry}

\newcommand{\FVBpm}{\ensuremath{\mathrm{FV}_{\PB^\pm}}}
\newcommand{\FVN}{\ensuremath{\mathrm{FV}_{\N}}}
\usepackage{microtype}
\usepackage{lineno}  % for line numbering during review
\usepackage{xspace} % To avoid problems with missing or double spaces after
\usepackage{caption} %these three command get the figure and table captions automatically small

\usepackage{graphicx}  % to include figures (can also use other packages)
\usepackage{color}
\usepackage{colortbl}
\graphicspath{{./figs/}} % Make Latex search fig subdir for figures
\usepackage{amsmath} % Adds a large collection of math symbols
\usepackage{amssymb}
\usepackage{amsfonts}
\usepackage{upgreek} % Adds in support for greek letters in roman typeset

\newcommand*\patchAmsMathEnvironmentForLineno[1]{%
\expandafter\let\csname old#1\expandafter\endcsname\csname #1\endcsname
\expandafter\let\csname oldend#1\expandafter\endcsname\csname
end#1\endcsname
 \renewenvironment{#1}%
   {\linenomath\csname old#1\endcsname}%
   {\csname oldend#1\endcsname\endlinenomath}%
}
\newcommand*\patchBothAmsMathEnvironmentsForLineno[1]{%
  \patchAmsMathEnvironmentForLineno{#1}%
  \patchAmsMathEnvironmentForLineno{#1*}%
}
\AtBeginDocument{%
\patchBothAmsMathEnvironmentsForLineno{equation}%
\patchBothAmsMathEnvironmentsForLineno{align}%
\patchBothAmsMathEnvironmentsForLineno{flalign}%
\patchBothAmsMathEnvironmentsForLineno{alignat}%
\patchBothAmsMathEnvironmentsForLineno{gather}%
\patchBothAmsMathEnvironmentsForLineno{multline}%
\patchBothAmsMathEnvironmentsForLineno{eqnarray}%
}

\usepackage[pdftex,
            pdfauthor={\paperauthors},
            pdftitle={\paperasciititle},
            pdfkeywords={\paperkeywords}]{hyperref}
\usepackage{hyperxmp}
\hypersetup{
    pdfcopyright={Copyright (C) \papercopyright},
    pdflicenseurl={\paperlicenceurl}
}
\usepackage[bottom,flushmargin,hang,multiple]{footmisc}

\usepackage[all]{hypcap} % Internal hyperlinks to floats.

\usepackage{xspace}
\usepackage{upgreek}

\def\BToNpi {\decay{\B}{\muon\N(\to \muon\pion)}}
\def\BToNpiX {\decay{\B}{\muon\N(\to \muon\pion) (\PX)}}
\def\NTomupi {\decay{\N}{\muon\pion}}

\def\lhcb   {\mbox{LHCb}\xspace}
\def\atlas  {\mbox{ATLAS}\xspace}
\def\cms    {\mbox{CMS}\xspace}

\def\babar  {\mbox{BaBar}\xspace}
\def\belle  {\mbox{Belle}\xspace}

\def\lhc    {\mbox{LHC}\xspace}

\def\bfactories {\mbox{\B Factories}\xspace}

\def\MagUp {\mbox{\em Mag\kern -0.05em Up}\xspace}

\ifthenelse{\boolean{uprightparticles}}%
{

 \def\Pmu         {\ensuremath{\upmu}\xspace}

 \def\Ppi         {\ensuremath{\uppi}\xspace}
 
 \def\Prho        {\ensuremath{\uprho}\xspace}
 
 \def\Ptau        {\ensuremath{\uptau}\xspace}

 \def\Ppsi        {\ensuremath{\uppsi}\xspace}

 \def\PDelta      {\ensuremath{\Delta}\xspace}
 \def\PXi         {\ensuremath{\Xi}\xspace}
 \def\PLambda     {\ensuremath{\Lambda}\xspace}
 \def\PSigma      {\ensuremath{\Sigma}\xspace}
 \def\POmega      {\ensuremath{\Omega}\xspace}
 \def\PUpsilon    {\ensuremath{\Upsilon}\xspace}
 \let\oldPi\Pi
 \def\PPi         {\ensuremath{\oldPi}\xspace}

 \def\PB      {\ensuremath{\mathrm{B}}\xspace}
 \def\PD      {\ensuremath{\mathrm{D}}\xspace}
 \def\PJ      {\ensuremath{\mathrm{J}}\xspace}
 \def\PK      {\ensuremath{\mathrm{K}}\xspace}
 \def\PN      {\ensuremath{\mathrm{N}}\xspace}
 \def\PW      {\ensuremath{\mathrm{W}}\xspace}
 \def\PX      {\ensuremath{\mathrm{X}}\xspace}
 \def\PZ      {\ensuremath{\mathrm{Z}}\xspace}
 \def\Pb      {\ensuremath{\mathrm{b}}\xspace}
 \def\Pc      {\ensuremath{\mathrm{c}}\xspace}
 \def\Pd      {\ensuremath{\mathrm{d}}\xspace}
 \def\Pe      {\ensuremath{\mathrm{e}}\xspace}
 \def\Pp      {\ensuremath{\mathrm{p}}\xspace}

 \def\Ps      {\ensuremath{\mathrm{s}}\xspace}
 
 \def\Pu      {\ensuremath{\mathrm{u}}\xspace}
 \def\thebaroffset{0.0em}
}
{

 \def\Pmu         {\ensuremath{\mu}\xspace}

 \def\Ppi         {\ensuremath{\pi}\xspace}
 
 \def\Prho        {\ensuremath{\rho}\xspace}
 
 \def\Ptau        {\ensuremath{\tau}\xspace}

 \def\Ppsi        {\ensuremath{\psi}\xspace}
 
 \mathchardef\PDelta="7101
 \mathchardef\PXi="7104
 \mathchardef\PLambda="7103
 \mathchardef\PSigma="7106
 \mathchardef\POmega="710A
 \mathchardef\PUpsilon="7107
 \mathchardef\PPi="7105
 \def\PB      {\ensuremath{B}\xspace}
 \def\PD      {\ensuremath{D}\xspace}
 \def\PJ      {\ensuremath{J}\xspace}
 \def\PK      {\ensuremath{K}\xspace}
 \def\PW      {\ensuremath{W}\xspace}
 \def\PZ      {\ensuremath{Z}\xspace}
 \def\Pb      {\ensuremath{b}\xspace}
 \def\Pc      {\ensuremath{c}\xspace}
 \def\Pd      {\ensuremath{d}\xspace}
 \def\Pe      {\ensuremath{e}\xspace}
 \def\Pp      {\ensuremath{p}\xspace}

 \def\Ps      {\ensuremath{s}\xspace}
 
 \def\Pu      {\ensuremath{u}\xspace}
 \def\thebaroffset{0.18em}
}
\newcommand{\offsetoverline}[2][\thebaroffset]{\kern #1\overline{\kern -#1 #2}}%

\makeatletter
\ifcase \@ptsize \relax% 10pt
  \newcommand{\miniscule}{\@setfontsize\miniscule{4}{5}}% \tiny: 5/6
\or% 11pt
  \newcommand{\miniscule}{\@setfontsize\miniscule{5}{6}}% \tiny: 6/7
\or% 12pt
  \newcommand{\miniscule}{\@setfontsize\miniscule{5}{6}}% \tiny: 6/7
\fi
\makeatother

\DeclareRobustCommand{\optbar}[1]{\shortstack{{\miniscule (\rule[.5ex]{1.25em}{.18mm})}
  \\ [-.7ex] $#1$}}

\def\electron   {{\ensuremath{\Pe}}\xspace}
\def\muon       {{\ensuremath{\Pmu}}\xspace}
\def\mup        {{\ensuremath{\Pmu^+}}\xspace}
\def\mun        {{\ensuremath{\Pmu^-}}\xspace} % muon negative (\mum is taken)
\def\mupm       {{\ensuremath{\Pmu^\pm}}\xspace}

\def\tauon      {{\ensuremath{\Ptau}}\xspace}

\def\N      {{\ensuremath{\PN}}\xspace}

\def\W      {{\ensuremath{\PW}}\xspace}

\def\Z      {{\ensuremath{\PZ}}\xspace}

\def\uquark    {{\ensuremath{\Pu}}\xspace}
\def\uquarkbar {{\ensuremath{\overline \uquark}}\xspace}

\def\dquark    {{\ensuremath{\Pd}}\xspace}

\def\squark    {{\ensuremath{\Ps}}\xspace}

\def\cquark    {{\ensuremath{\Pc}}\xspace}
\def\cquarkbar {{\ensuremath{\overline \cquark}}\xspace}

\def\bquark    {{\ensuremath{\Pb}}\xspace}
\def\bquarkbar {{\ensuremath{\overline \bquark}}\xspace}

\def\pion   {{\ensuremath{\Ppi}}\xspace}

\def\pip    {{\ensuremath{\pion^+}}\xspace}
\def\pim    {{\ensuremath{\pion^-}}\xspace}

\def\pimp   {{\ensuremath{\pion^\mp}}\xspace}

\def\kaon    {{\ensuremath{\PK}}\xspace}
\def\KorKbar {\kern \thebaroffset\optbar{\kern -\thebaroffset \PK}{}\xspace}

\def\Km      {{\ensuremath{\kaon^-}}\xspace}

\def\D       {{\ensuremath{\PD}}\xspace}

\def\DorDbar {\kern \thebaroffset\optbar{\kern -\thebaroffset \PD}\xspace}
\def\Dz      {{\ensuremath{\D^0}}\xspace}

\def\Dp      {{\ensuremath{\D^+}}\xspace}
\def\Dm      {{\ensuremath{\D^-}}\xspace}

\def\DpDm    {\ensuremath{\Dp {\kern -0.16em \Dm}}\xspace}

\def\Dstarp  {{\ensuremath{\D^{*+}}}\xspace}

\def\B       {{\ensuremath{\PB}}\xspace}

\def\BorBbar {\kern \thebaroffset\optbar{\kern -\thebaroffset \PB}\xspace}
\def\Bz      {{\ensuremath{\B^0}}\xspace}

\def\Bd      {{\ensuremath{\B^0}}\xspace}

\def\BdorBdbar {\kern \thebaroffset\optbar{\kern -\thebaroffset \Bd}\xspace}
\def\Bu      {{\ensuremath{\B^+}}\xspace}

\def\Bp      {{\ensuremath{\Bu}}\xspace}

\def\Bpm     {{\ensuremath{\B^\pm}}\xspace}

\def\Bs      {{\ensuremath{\B^0_\squark}}\xspace}

\def\BsorBsbar {\kern \thebaroffset\optbar{\kern -\thebaroffset \Bs}\xspace}
\def\Bc      {{\ensuremath{\B_\cquark^+}}\xspace}
\def\Bcp     {{\ensuremath{\B_\cquark^+}}\xspace}

\def\jpsi     {{\ensuremath{{\PJ\mskip -3mu/\mskip -2mu\Ppsi}}}\xspace}

\def\Y#1S{\ensuremath{\PUpsilon{(#1S)}}\xspace}

\def\proton      {{\ensuremath{\Pp}}\xspace}

\def\LorLbar     {\kern \thebaroffset\optbar{\kern -\thebaroffset \PLambda}\xspace}

\newcommand{\decay}[2]{\mbox{\ensuremath{#1\!\to #2}}\xspace}

\def\to                 {\ensuremath{\rightarrow}\xspace}

\def\Vub  {{\ensuremath{V_{\uquark\bquark}^{\phantom{\ast}}}}\xspace}
\def\Vcb  {{\ensuremath{V_{\cquark\bquark}^{\phantom{\ast}}}}\xspace}

\def\AT#1     {\ensuremath{A_{\mathrm{T}}^{#1}}\xspace}           % 2

\def\C#1      {\ensuremath{\mathcal{C}_{#1}}\xspace}                       % 9
\def\Cp#1     {\ensuremath{\mathcal{C}_{#1}^{'}}\xspace}                    % 7
\def\Ceff#1   {\ensuremath{\mathcal{C}_{#1}^{\mathrm{(eff)}}}\xspace}        % 9  
\def\Cpeff#1  {\ensuremath{\mathcal{C}_{#1}^{'\mathrm{(eff)}}}\xspace}       % 7
\def\Ope#1    {\ensuremath{\mathcal{O}_{#1}}\xspace}                       % 2
\def\Opep#1   {\ensuremath{\mathcal{O}_{#1}^{'}}\xspace}                    % 7

\newcommand{\nospaceunit}[1]{\ensuremath{\text{#1}}}
\newcommand{\aunit}[1]{\ensuremath{\text{\,#1}}}
\newcommand{\tev}{\aunit{Te\kern -0.1em V}\xspace}
\newcommand{\gev}{\aunit{Ge\kern -0.1em V}\xspace}
\newcommand{\mev}{\aunit{Me\kern -0.1em V}\xspace}
\newcommand{\kev}{\aunit{ke\kern -0.1em V}\xspace}
\newcommand{\ev}{\aunit{e\kern -0.1em V}\xspace}

\newcommand{\mevc}{\ensuremath{\aunit{Me\kern -0.1em V\!/}c}\xspace}
\newcommand{\gevc}{\ensuremath{\aunit{Ge\kern -0.1em V\!/}c}\xspace}
\newcommand{\mevcc}{\ensuremath{\aunit{Me\kern -0.1em V\!/}c^2}\xspace}
\newcommand{\gevcc}{\ensuremath{\aunit{Ge\kern -0.1em V\!/}c^2}\xspace}
\def\m    {\aunit{m}\xspace}

\def\cm   {\aunit{cm}\xspace}

\def\mm   {\aunit{mm}\xspace}

\def\mum  {\ensuremath{\,\upmu\nospaceunit{m}}\xspace}

\def\mub{\ensuremath{\,\upmu\nospaceunit{b}}\xspace}

\def\fb   {\ensuremath{\aunit{fb}}\xspace}
\def\invfb   {\ensuremath{\fb^{-1}}\xspace}

\def\ns   {\ensuremath{\aunit{ns}}\xspace}
\def\ps   {\ensuremath{\aunit{ps}}\xspace}

\newcommand{\chisq}{\ensuremath{\chi^2}\xspace}

\newcommand{\chisqip}{\ensuremath{\chi^2_{\text{IP}}}\xspace}

\def\gsim{{~\raise.15em\hbox{$>$}\kern-.85em
          \lower.35em\hbox{$\sim$}~}\xspace}
\def\lsim{{~\raise.15em\hbox{$<$}\kern-.85em
          \lower.35em\hbox{$\sim$}~}\xspace}

\def\sqs   {\ensuremath{\protect\sqrt{s}}\xspace}

\def\pt         {\ensuremath{p_{\mathrm{T}}}\xspace}

\def\ptot       {\ensuremath{p}\xspace}

\def\evtgen     {\mbox{\textsc{EvtGen}}\xspace}

\def\geant      {\mbox{\textsc{Geant4}}\xspace}

\def\photos     {\mbox{\textsc{Photos}}\xspace}

\def\pythia     {\mbox{\textsc{Pythia}}\xspace}

\def\tell1  {TELL1\xspace}
\def\ukl1   {UKL1\xspace}

\newcommand{\eg}{\mbox{\itshape e.g.}\xspace}
\newcommand{\ie}{\mbox{\itshape i.e.}\xspace}

\newcommand{\etc}{\mbox{\itshape etc.}\xspace}

\newcommand{\lhcborcid}[1]{\href{https://orcid.org/#1}{\hspace*{0.1em}\raisebox{-0.45ex}{\includegraphics[width=1em]{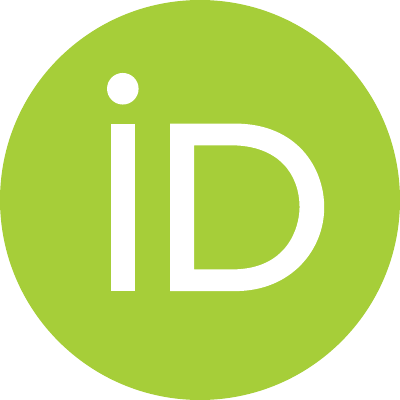}}}}

\hypersetup{
  colorlinks   = true, %Colours links instead of ugly boxes
  urlcolor     = blue, %Colour for external hyperlinks
  linkcolor    = blue, %Colour of internal links
  citecolor    = red   %Colour of citations
}

\ifEnableSectionTOCLinks
    \usepackage[explicit]{titlesec} % to change headings
    
    \let\oldcontentsline\contentsline
    \renewcommand

    \titleformat{\section}{\normalfont\Large\bf}{\hyperlink{tocsection.\thesection}{{\thesection} \parbox[t]{\dimexpr\textwidth-1pc}{#1}}}{1pc}{}

    \titleformat{\subsection}{\normalfont\bf}{\hyperlink{tocsubsection.\thesubsection}{{\thesubsection} \parbox[t]{\dimexpr\textwidth-1pc}{#1}}}{1pc}{}

    \titleformat{name=\section,numberless}[display]{}{}{0pt}{\normalfont\Huge\bfseries #1}
\fi

\usepackage{cite} % Allows for ranges in citations
\usepackage{mciteplus}
\usepackage{longtable} % only for template; not usually to be used in PAPERs

\begin{document}

%%%%%%%%%%%%%%%%%%%%%%%%%
%%%%% Title     %%%%%%%%%
%%%%%%%%%%%%%%%%%%%%%%%%%
\renewcommand{\thefootnote}{\fnsymbol{footnote}}
\setcounter{footnote}{1}

% %%%%%%% CHOOSE TITLE PAGE--------
%\onecolumn
%\input{title-LHCb-INT}
%\input{title-LHCb-ANA}
%\input{title-LHCb-CONF}
%\input{title-LHCb-FIGURE}
% ===============================================================================
% Purpose: LHCb-PAPER journal paper title page template
% Author: 
% Created on: 2010-09-25
% ===============================================================================

%%%%%%%%%%%%%%%%%%%%%%%%%
%%%%%  TITLE PAGE  %%%%%%
%%%%%%%%%%%%%%%%%%%%%%%%%
\begin{titlepage}
\pagenumbering{roman}

% Header ---------------------------------------------------
\vspace*{-1.5cm}
\centerline{\large EUROPEAN ORGANIZATION FOR NUCLEAR RESEARCH (CERN)}
\vspace*{1.5cm}
\noindent
\begin{tabular*}{\linewidth}{lc@{\extracolsep{\fill}}r@{\extracolsep{0pt}}}
\ifthenelse{\boolean{pdflatex}}% Logo format choice
{\vspace*{-1.5cm}\mbox{\!\!\!\includegraphics[width=.14\textwidth]{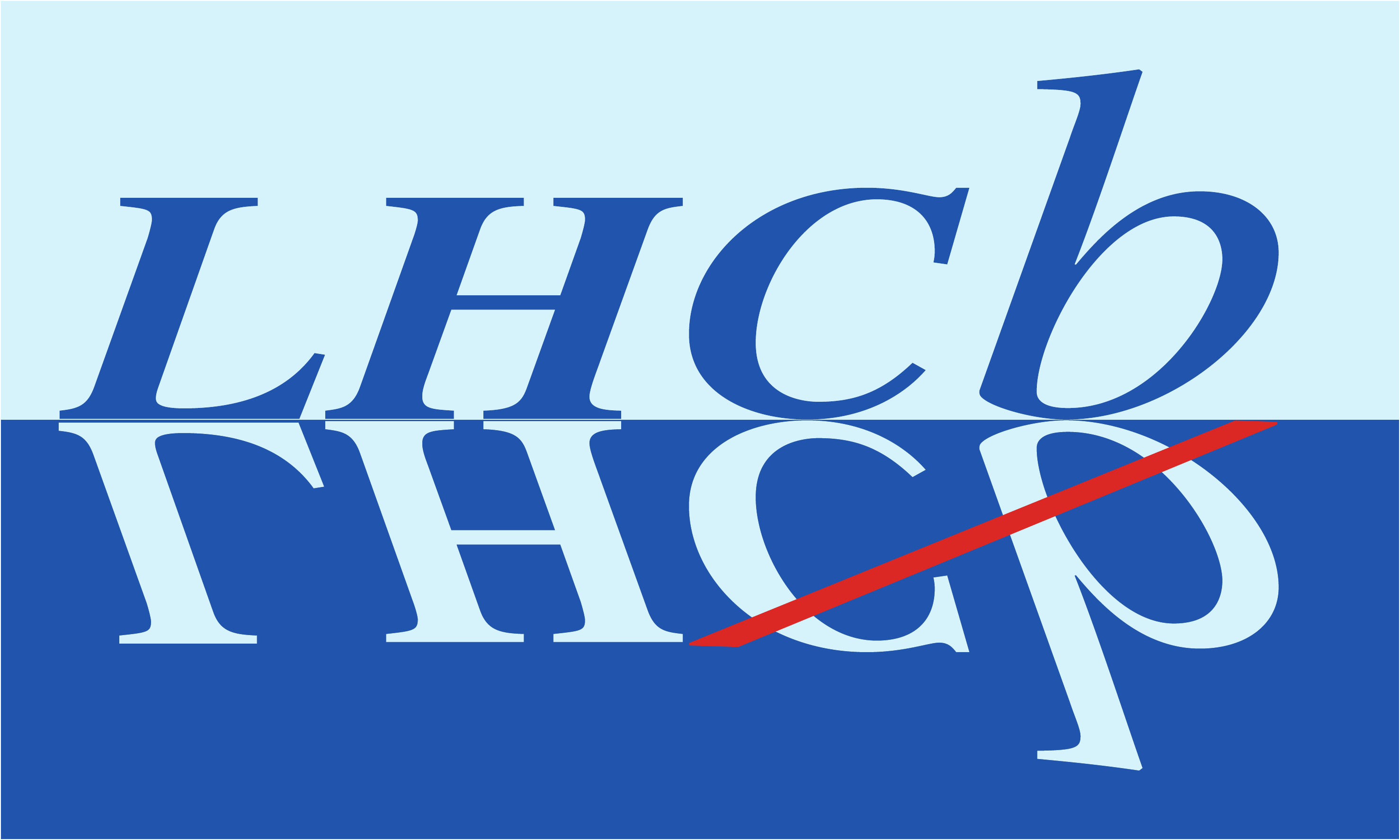}} & &}%
{\vspace*{-1.2cm}\mbox{\!\!\!\includegraphics[width=.12\textwidth]{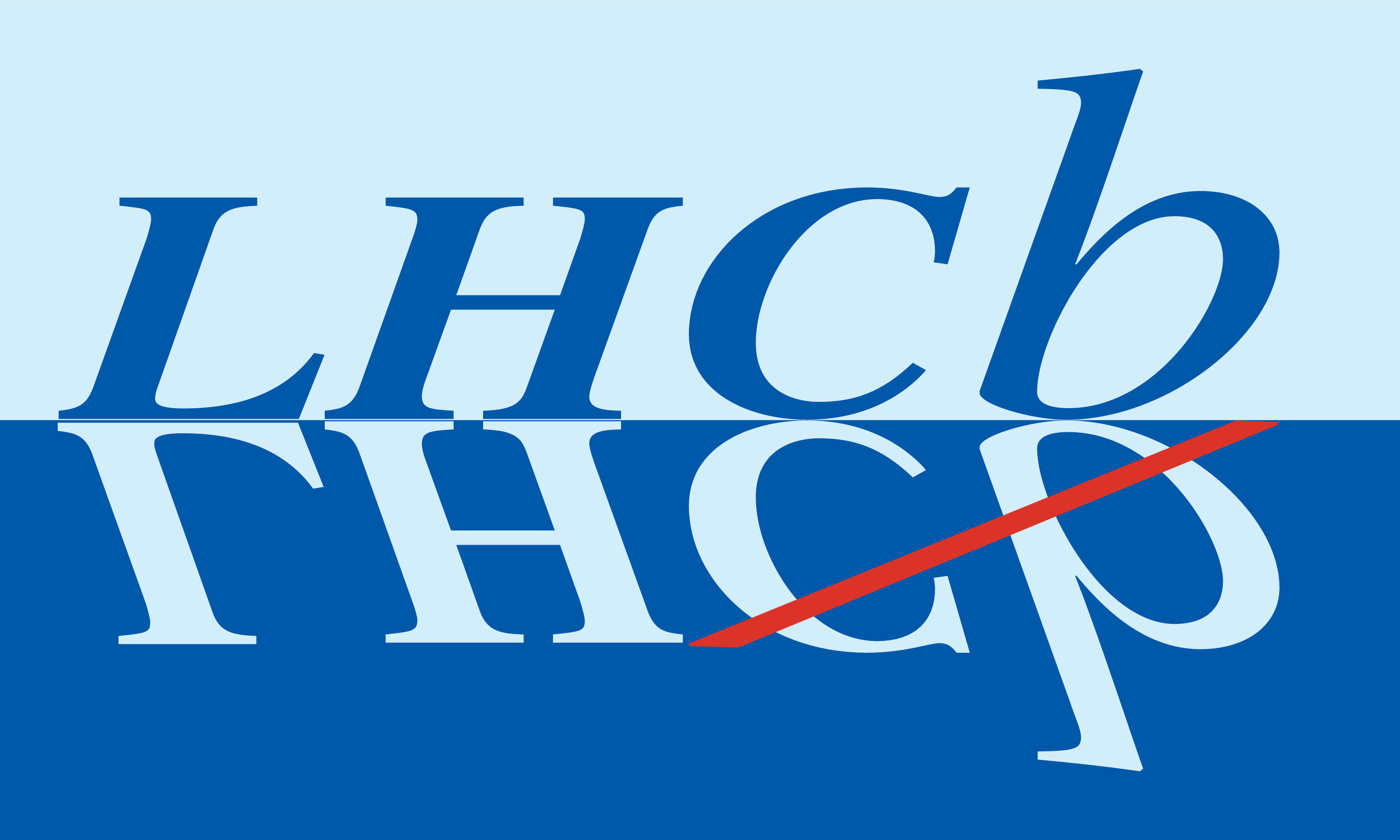}} & &}%
\\
 & & CERN-EP-2025-264 \\  % ID 
 & & LHCb-PAPER-2025-042 \\  % ID 
 & & March 30, 2026 \\%\today \\ % Date - Can also hardwire e.g.: 23 March 2010
 & & \\
% not in paper \hline
\end{tabular*}

\vspace*{4.0cm}

% Title --------------------------------------------------
{\normalfont\bfseries\boldmath\huge
\begin{center}
% DO NOT EDIT HERE. Instead edit macro in main.tex to keep metadata correct
  \papertitle 
\end{center}
}

\vspace*{2.0cm}

% Authors -------------------------------------------------
\begin{center}
%In the footnote, replace 'paper' by 'Letter' in case of submission to PRL or PLB 
% Edit macro in main.tex to keep metadata correct
\paperauthors\footnote{Authors are listed at the end of this paper.}
\end{center}

\vspace{\fill}

% Abstract -----------------------------------------------
\begin{abstract}
\noindent
A search for long-lived heavy neutral leptons produced in B-meson decays and decaying to a $\upmu^\pm\uppi^\mp$ final state is performed with data collected by the LHCb experiment in proton--proton collisions at a centre-of-mass energy of 13~TeV, corresponding to an
integrated luminosity of $5\,\mathrm{fb}^{-1}$.
The results are interpreted in both lepton-number-conserving and lepton-number-violating scenarios. No significant excess is observed. Constraints are placed on the squared mixing element $|U_{\muon \N}|^2$ to the active muon neutrino, under the assumption that couplings to other lepton flavours are negligible, in the mass range of $1.6$--$5.5$~GeV.
\end{abstract}

\vspace*{2.0cm}

\begin{center}
  Published in Journal of High Energy Physics 03 (2026) 178 %/
 % Phys.~Rev.~D /
 % Eur.~Phys.~J.~C 
\end{center}

\vspace{\fill}

{\footnotesize 
% Edit macro in main.tex to keep metadata correct
\centerline{\copyright~\papercopyright. \href{\paperlicenceurl}{\paperlicence}.}}
\vspace*{2mm}

\end{titlepage}

%%%%%%%%%%%%%%%%%%%%%%%%%%%%%%%%
%%%%%  EOD OF TITLE PAGE  %%%%%%
%%%%%%%%%%%%%%%%%%%%%%%%%%%%%%%%

%  empty page follows the title page ----
\newpage
\setcounter{page}{2}
\mbox{~}
%\newpage
%
%% Author List ----------------------------
%%  You need to get a new author list!
%\input{LHCb_authorlist.tex}
%
%The author list for journal publications is provided by the Membership Committee shortly after 'approval to go to paper' has been given.
%%It will be made available on the page
%%\verb!http://www.physik.uzh.ch/~strauman/forMemCo/LHCb-PAPER-XXXX-XXX/! .
%It will be sent to you by email shortly after a paper number has beens assigned.
%The author list should be included already at first circulation, 
%to allow new members of the collaboration to verify whether they have been included correctly.
%Occasionally a misspelled name is corrected or associated institutions become full members.
%In that case, a new author list will be sent to you.
%In case line numbering doesn't work well after including the authorlist, try moving the \verb!\bigskip! after the last author to a separate line.
%
%
%The authorship for Conference Reports should be ``The LHCb
%  collaboration'', with a footnote giving the name(s) of the contact
%  author(s), but without the full list of collaboration names.

%\twocolumn
% %%%%%%%%%%%%% ---------

\renewcommand{\thefootnote}{\arabic{footnote}}
\setcounter{footnote}{0}

%%%%%%%%%%%%%%%%%%%%%%%%%%%%%%%%
%%%%%  Table of Content   %%%%%%
%%%%%%%%%%%%%%%%%%%%%%%%%%%%%%%%
%%%% Uncomment if desired
%\tableofcontents

\cleardoublepage

%%%%%%%%%%%%%%%%%%%%%%%%%
%%%%% Main text %%%%%%%%%
%%%%%%%%%%%%%%%%%%%%%%%%%

\pagestyle{plain} % restore page numbers for the main text
\setcounter{page}{1}
\pagenumbering{arabic}

%% Uncomment during review phase. 
%% Comment before a final submission.
%\linenumbers

%% This is the main body
%% It is useful to have a single file so comments are not missed in overleaf.
\section{Introduction}
\label{sec:Introduction}

The Standard Model (SM) of particle physics, despite its remarkable success in describing the fundamental forces and particles of the Universe, leaves several crucial questions unanswered, including: the nature of dark matter~\cite{Bertone:2010zza}, the matter-antimatter asymmetry in the Universe~\cite{Farrar:1993hn}, and the origin of neutrino masses~\cite{Bilenky:2016pep}. These gaps in the current understanding suggest the existence of new physics beyond the SM, already established by the discovery of neutrino oscillations~\cite{Super-Kamiokande:1998kpq}, which provided the experimental evidence for nonzero neutrino masses.

Heavy neutral leptons (HNLs, denoted by \N) are hypothetical fermions that are singlets under all SM gauge groups and appear as right-handed neutrinos in type-I seesaw extensions of the SM~\cite{Minkowski:1977sc,Asaka:2005an,Asaka:2005pn}. 
In these frameworks, the neutrino sector contains both Dirac and Majorana mass terms, with the latter enabling lepton-number violation.
The mixing between HNLs and active neutrinos depends on the ratio of Dirac to Majorana mass scales, inducing suppressed couplings $U_{\ell\N}$ ($\ell=\electron,\muon,\tauon$) that permit HNL production in weak decays of hadrons and gauge bosons, and subsequent decays back into SM particles via the same interaction~\cite{King:2004,Altarelli:2004}. 
If the HNL is a Majorana particle, lepton-number-violating signatures such as same-sign dimuons are possible.

The Minimal Neutrino Standard Model ($\nu$MSM) provides a benchmark framework, augmenting the SM with three singlet HNL states~\cite{Asaka:2005an,Asaka:2005pn}. In this model, two $\mathcal{O}(\!\gev)$ mass\footnote{Natural units with $\hbar=c=1$ are used throughout.} HNLs can generate the needed amount of baryon asymmetry via low-scale leptogenesis, while an $\mathcal{O}(\!\kev)$ state can account for dark matter through resonant production~\cite{Fukugita:1986hr,Canetti:2012kh}. However, successful leptogenesis in the $\nu$MSM typically requires total mixing \mbox{$U^2 \equiv \sum_{\ell} |U_{\ell \N}|^2 \lesssim 10^{-6}$} for HNL masses, $m_\N$, of $\mathcal{O}(\!\gev)$~\cite{Klaric:2021cpi,Hernandez:2022ivz}, below the sensitivity of this analysis. In contrast, scenarios with more HNLs or additional interactions, such as models with symmetry-protected low-scale seesaw mechanisms (\eg an inverse seesaw with approximate lepton-number symmetry), flavour-structured models with dominant muon mixing, or frameworks with gauged difference between baryon and lepton number predict 
\mbox{$|U_{\muon \N}|^2 \sim 10^{-5}$--$10^{-4}$} for \gev-scale HNLs while remaining consistent with light-neutrino masses and cosmological constraints~\cite{Drewes:2021nqr,Drewes:2024bla,ShivaSankar:2025}. This search is therefore formulated in a model-agnostic way in the $(m_\N, |U_{\muon \N}|^2)$ plane, targeting the large-mixing regime relevant to these scenarios.
The lifetime of a given HNL, $\tau_\N$, depends on the HNL mass and the elements of the mixing matrix. In particular, the lifetime varies as $\tau_\N\sim |U_{\ell \N}|^{-2} m_\N^{-5}$~\cite{Asaka:2005an}. Hence, for sufficiently low values of mixing-matrix elements, even HNLs in the \gev--\tev range can be long-lived ($\tau_\N \gtrsim 1\ps$).

Searches for HNLs have been performed by several experiments using a comprehensive set of methodologies. The most stringent constraints on the mixing-matrix elements for HNLs with masses up to about 2\gev were set by beam-dump experiments looking at HNLs produced from \D and \kaon meson decays, \eg by the PS191~\cite{Bernardi:1987ek}, E949~\cite{E949:2014gsn}, CHARM~\cite{DORENBOSCH1986473}, NuTeV~\cite{NuTeV:1999HNLs} and BEBC~\cite{BEBC:1985HNLs} experiments. Searches at \bfactories were conducted by the \babar~\cite{BaBar:2022cqj} and \belle~\cite{Liventsev:2013zz,Belle:2022tfo} experiments, searching for HNLs produced by \B mesons ($m_\N \lesssim 6$\gev). At the CERN Large Hadron Collider (\lhc), limits on heavy neutral leptons have been set by the \cms~\cite{CMS:2018iaf,CMS:2018jxx,CMS:2022fut, CMS:2024HNLmuondetector, CMS:2024HNLInclusive,CMS:2024xdq,CMS:2024bni}, \atlas~\cite{Aad:2019kiz,ATLAS:2022atq} and LHCb~\cite{LHCb-PAPER-2013-064,LHCb-PAPER-2020-022} experiments on HNLs produced from \B mesons, as well as from \W and \Z bosons (either real or virtual).

\begin{figure}[tb]
  \centering
  \includegraphics{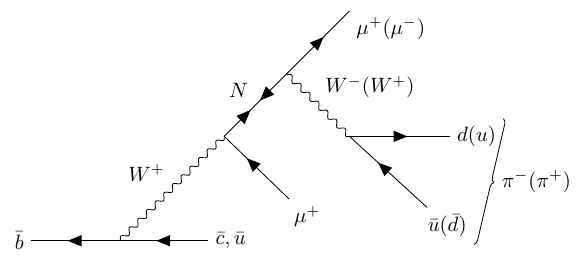}
  \caption{
  Feynman diagram for the inclusive HNL production from \bquarkbar quarks and its subsequent decay $\N\to\mupm\pimp$. This diagram represents various decay channels depending on the hadronisation of the \bquarkbar quark. The \bquarkbar quark and other quarks (\cquarkbar or \uquarkbar) can either form the initial-state meson or appear in the final state. If the HNL is of a Majorana nature, the two muons in the final state can have the same sign.
  }
  \label{fig:feynman}
\end{figure}

The search presented in this paper targets long-lived HNLs produced in \B-meson decays and with lifetimes corresponding to a significant decay probability within the LHCb detector's acceptance. 
The analysis is based on proton-proton ($\proton\proton$) collision data recorded at a centre-of-mass energy of $13\tev$ with the \lhcb detector during 2016--2018 (Run~2), corresponding to an integrated luminosity of ${\cal L}=5.04\pm0.10\,\invfb$.
A search is performed for HNLs produced in both fully reconstructed leptonic $\decay{\B^+_{(\cquark)}}{\mup \N}$ and partially reconstructed semileptonic $\decay{\B}{\mup \N \PX}$ decays, where $\PX$ represents additional unreconstructed particles, and where the HNL decays to $\mup \pim$.\footnote{The inclusion of charge-conjugate processes is implied throughout.} The search takes the form of a bump hunt in the invariant mass of the HNL decay products. A Feynman diagram for the searched HNL production and decay modes is shown in Fig.~\ref{fig:feynman}.
The muons from the \B meson and \N decays may have the same sign, enabling sensitivity to lepton-number-violating processes induced by a Majorana-type mass term. Majorana- and Dirac-like (\ie lepton-number violating and conserving) cases are treated independently in the search. 
The search is sensitive only to HNLs that mix with muon neutrinos, and therefore the results are interpreted in terms of the $|U_{\muon \N}|^2$ mixing-matrix element. Couplings with other lepton species are assumed to be negligible, that is: $U_{\electron \N}=U_{\tauon \N}=0$.

The search focuses on an HNL mass range extending from just above the charm-meson mass, up to the mass of the \Bc meson.
For HNLs in this mass range, \B-meson decays are the dominant HNL production mode at the \lhc~\cite{Bondarenko:2018ptm}.
The production of HNLs from beauty baryons has been calculated in Ref.~\cite{Ramazanov:2008ph} and is expected to be suppressed relative to production from beauty mesons~\cite{Bondarenko:2018ptm}. Baryonic production is therefore neglected in this analysis. Some small peripheral sensitivity contribution is nonetheless expected, but is not considered in the computed limits, making them somewhat conservative. Although the \bquark-quark fragmentation fraction to \Bc mesons is small at the \lhc~\cite{LHCb-PAPER-2019-033}, HNL production from leptonic \Bc decays is enhanced by the ratio of the CKM matrix elements $|\Vcb|^2/|\Vub|^2$ compared to production from leptonic \Bp decays, and thus makes an important contribution to the analysis.

\section{\lhcb detector, trigger and simulation}

The \lhcb detector~\cite{LHCb-DP-2008-001,LHCb-DP-2014-002} is a single-arm forward spectrometer covering the \mbox{pseudorapidity} range $2<\eta <5$, designed for the study of particles containing \bquark or \cquark quarks. The detector used for this analysis includes a high-precision tracking system consisting of a silicon-strip vertex detector (VELO) surrounding the $\proton\proton$ interaction region~\cite{LHCb-DP-2014-001}, a large-area silicon-strip detector (TT) located upstream of a dipole magnet with a bending power of about $4{\mathrm{\,T\,m}}$, and three stations of silicon-strip detectors and straw drift tubes (IT and OT)~\cite{LHCb-DP-2017-001} placed downstream of the magnet. 
Different track types are reconstructed based on the subdetectors registering charged particle hits.
Tracks formed with the track segments in the vertex detector, upstream and downstream tracking stations are referred to as \emph{long} tracks, while those having hits only in the upstream and downstream tracking stations are called \emph{downstream} tracks. Compared to downstream tracks, the long tracks have better momentum, \ptot, and impact parameter, IP, resolution relative to the primary \proton\proton collision, PV. For long tracks,
the momentum resolution varies from 0.5\% at low momentum to 1.0\% at 200\gev, while the impact parameter resolution is
 $(15+29/\pt)\mum$, where \pt is the component of the momentum transverse to the beam, in\,\gev.

Different types of charged hadrons are distinguished using information from two ring-imaging Cherenkov detectors~\cite{LHCb-DP-2012-003}, providing particle identification (PID). Muons are identified by a system composed of alternating layers of iron and multiwire proportional chambers~\cite{LHCb-DP-2012-002}. 
Photons, electrons and hadrons are identified by a calorimeter system
consisting of scintillating-pad (SPD) and preshower (PS) detectors, an electromagnetic and a hadronic calorimeter~\cite{LHCb-DP-2020-001}. The SPD identifies charged particles, aids in separating electrons from photons, and measures event multiplicity through the number of hits ($n_{\rm SPD}$).

The online event selection is performed by a trigger~\cite{LHCb-DP-2012-004}, which consists of a hardware stage, based on information from the calorimeter and muon systems, followed by a software stage, which applies a full event reconstruction. The variable $n_{\rm SPD}$ is used in the hardware trigger stage (L0) for a global event cut (GEC). Specifically, the single-muon lines apply $n_{\rm SPD}<450$ and dimuon lines $n_{\rm SPD}<900$. To harmonise efficiencies, an offline $n_{\rm SPD}<450$ requirement is imposed on all events. Triggered data further undergo a centralised, offline processing step~\cite{Stripping}.

Simulation is used to model the effects of the detector acceptance and the imposed selection requirements. In the simulation, $\proton\proton$ collisions are generated using \pythia~\cite{Sjostrand:2007gs,Sjostrand:2006za}  with a specific \lhcb configuration~\cite{LHCb-PROC-2010-056}. Decays of unstable particles are described by \evtgen~\cite{Lange:2001uf}, in which final-state radiation is generated using \photos~\cite{davidson2015photos}. The interaction of the generated particles with the detector, and its response, are implemented using the \geant toolkit~\cite{Allison:2006ve, Agostinelli:2002hh} as described in Ref.~\cite{LHCb-PROC-2011-006}.

\begin{figure}[b]
    \centering
    \includegraphics[width=.8\textwidth]{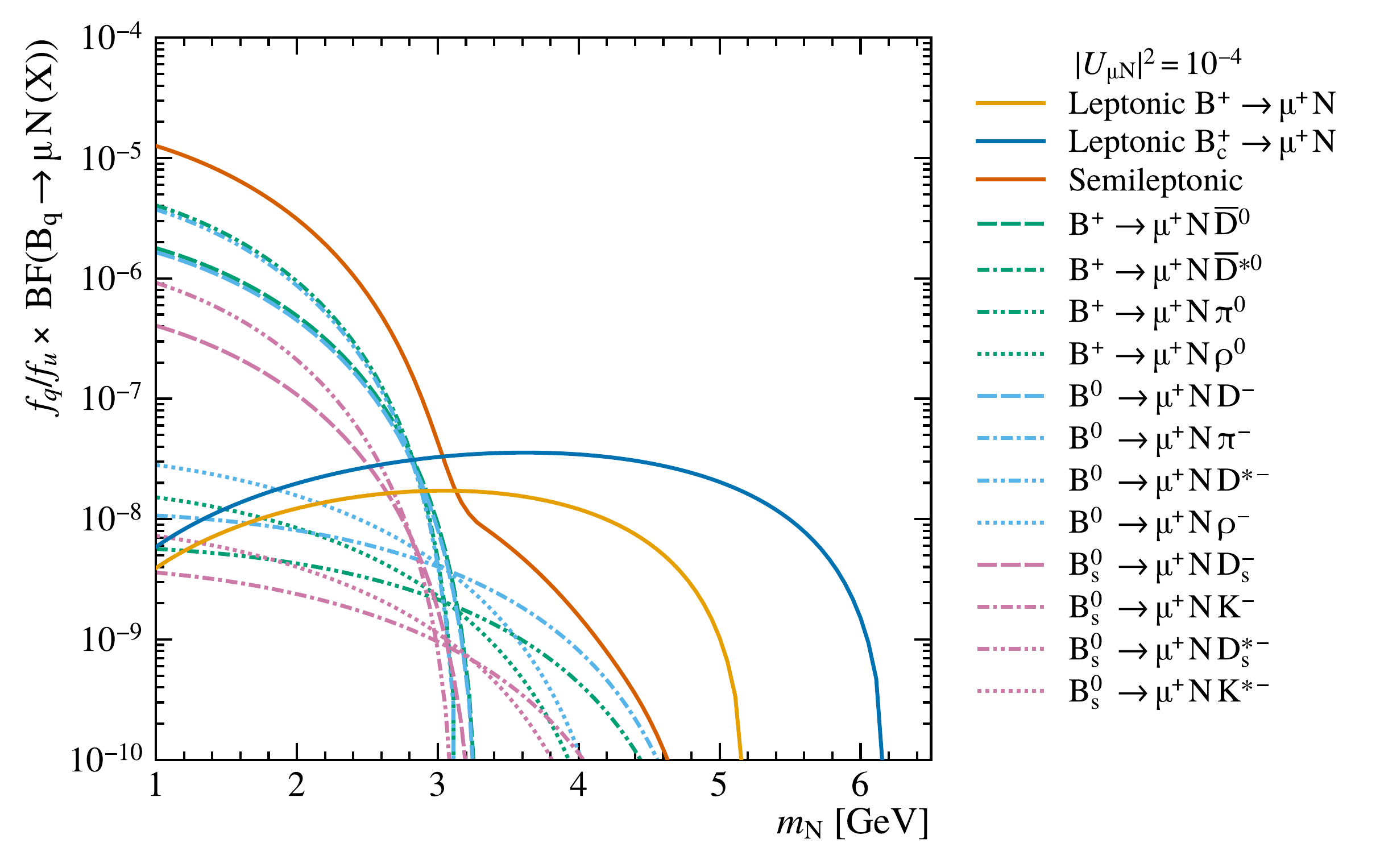}
    \caption{Branching fractions of various \B-meson channels corresponding to different \B-meson types, \Bz, \Bp, \Bs or \Bc, weighted by the corresponding fragmentation fraction. These values depend on the HNL mass, $m_\N$, and are calculated for a mixing parameter of $|U_{\muon \N}|^2=10^{-4}$. The solid red ``Semileptonic" line is the sum of all semileptonic decay channels, \ie the sum of all the specific decays below the semileptonic line in the legend.}
    \label{fig:BFs}
\end{figure}

Several simulated samples of HNL decays are produced considering different mass hypotheses, $m_\N = 1.6$, $2.0$, $3.0$, $4.0$, $5.0$ and $5.5\gev$, and lifetimes, $\tau_\N =10$, $100$ and $1000\ps$. 
This set of lifetimes has been chosen to cover the decay-time acceptance of the \lhcb detector for displaced HNL decays to \emph{long} or \emph{downstream} tracks. 
An independent sample is generated for each HNL mass, lifetime and \B-meson type: \Bz, \Bp, \Bs or \Bc. 
To model the total signal, the relevant mesons and decay modes are merged together taking into account the relative $\bquark$-quark fragmentation fractions to the different \B-meson species, $f_\squark/f_\uquark$ and $f_\cquark/f_\uquark$, measured by the \lhcb collaboration~\cite{LHCb-PAPER-2018-050,LHCb-PAPER-2019-033}, and the branching-ratio predictions of \B decays to HNLs taken from Ref.~\cite{Bondarenko:2018ptm}. 
The branching fractions of various decay channels, weighted by their corresponding fragmentation fractions, are presented in Fig.~\ref{fig:BFs}. These values depend on the HNL mass $m_\N$, and are calculated for a mixing parameter of $|U_{\muon \N}|^2=10^{-4}$.

Both the opposite sign (OS) lepton-number-conserving \mup\mun and same sign (SS) lepton-number-violating \mup\mup decay channels are simulated. All \B-meson decays involving HNLs and \decay{\N}{\muon\pion} decays are simulated neglecting HNL spin and polarisation. The same simplified model was used in previous searches of HNLs from \B decays at the \belle~\cite{Liventsev:2013zz}, \lhcb~\cite{LHCb-PAPER-2013-064}, and \cms~\cite{CMS:2024HNLInclusive} experiments.

\section{Event selection}
\label{sec:selection}
The flight distance of the HNL is a key observable for identifying displaced decays in the \lhcb detector. It is defined as $\mbox{\(\mathrm{FD}(\N) \equiv \sqrt{(x_{\N} - x_{\mathrm{B}})^2 + (y_{\N} - y_{\mathrm{B}})^2 + (z_{\N} - z_{\mathrm{B}})^2}\)}$, representing the spatial separation between the \B-meson and \N decay vertices. Owing to the forward geometry of the detector, the longitudinal component, $\mbox{\(\mathrm{FD}_z(\N) = z_{\N} - z_{\mathrm{B}}\)}$, provides an effective proxy for the full three-dimensional distance.

In each event, multiple $\proton\proton$ interactions are reconstructed. The PV that fits best to the flight direction of the \B candidate is taken as the PV associated to the \B decay.

Candidate \BToNpiX decays are separated into two analysis categories, referred to as \emph{long} and \emph{downstream}. The \emph{long} category includes \NTomupi decays that occur within the VELO and are reconstructed with \emph{long} tracks, which offer superior mass, momentum, and vertex resolution. In this category, candidates must have \(\mathrm{FD}_z(\N)\) greater than 20 mm. The \emph{downstream} category is composed of candidates that decay after the VELO and are reconstructed using \emph{downstream} tracks. This is sensitive to HNLs with ${\rm FD}_z(\N) \gtrsim 70\cm$.

The muon candidates from the signal decay are required to pass the muon-based L0 hardware trigger lines. These trigger lines accept events with either one or two muon candidates having a $\pt$ above specific thresholds.
For single muon candidates, the thresholds are $1.30\gev$ in 2016, $1.70\gev$ in 2017, and $1.75\gev$ in 2018.
For dimuon candidates, the product of their transverse momenta must exceed $1.69\gev^2$ in 2016 and $3.24\gev^2$ in 2017 and 2018.
 
In the subsequent software trigger, at least one of the muon candidates is required to have \mbox{$\pt>1.1\gev$} and $\ptot>6.0\gev$. The muon candidates that satisfy these momentum criteria are also required to have associated tracks with good track-fit quality and to be inconsistent with originating from any PV in the event.\footnote{The size of the impact parameter is measured relative to its uncertainty with the variable \chisqip, defined as the difference in the vertex-fit \chisq of a given PV reconstructed with and
without the particle being considered.} 

At the second software trigger stage, the \emph{long} category is required to pass either an inclusive topological selection targeting displaced \bquark-hadron vertices or an inclusive selection for displaced \NTomupi decays. The \emph{downstream} category is selected with a trigger specifically developed for this analysis that selects \BToNpi candidates with same-sign muons.
In this trigger, the pion and muon candidates that form the HNL candidate are required to have $\pt > 250\mev$ and a momentum exceeding 2\gev and 3\gev, respectively. The resulting HNL candidate must have $\pt > 700\mev$ and is combined with a long muon candidate to form a \B candidate. The reconstructed \B momentum is required to be aligned with the displacement vector from the associated PV to the \B secondary vertex, in the sense that the relative angle $\theta$ satisfies $\cos\theta > 0.99$. This trigger is prescaled by a factor of five, effectively analysing 20\% of these candidates.

In the offline selection, a \BToNpiX candidate must satisfy loose kinematic and PID requirements before being processed by a dedicated and bespoke multivariate analysis (MVA). The \N candidate is formed from two oppositely charged tracks, which must be either both long tracks or both downstream tracks. These tracks, identified as a pion candidate and a muon candidate, are required to pass the same momentum and transverse momentum selections as in the downstream trigger for the second software stage, and to have a large impact parameter with respect to the associated PV, selecting those consistent with a long-lived \N and rejecting prompt tracks from the primary $\proton\proton$ interaction.
The two tracks are required to originate from a common vertex with good fit \chisq,  and the vertex is required to be significantly displaced from the associated PV, as quantified by \(\mathrm{FD}_z(\N)\).
The \N candidate with $\pt > 700\mev$ and $m_{\N} > 1.5\gev$ is combined with a long \muon candidate to form a \B candidate, requiring a common vertex with good fit \chisq. The \B momentum is required to pass the same alignment criteria with the associated PV as in the downstream trigger. The mass of the \B candidate, $m(\muon\muon\pion)$, should be in the wide range $1.5<m(\muon\muon\pion)<6.5\gev$. These requirements are sufficiently loose to still select partially reconstructed \BToNpiX decays efficiently. 

A fiducial volume (FV) selection removes candidates in low-acceptance regions, requiring final-state tracks with $2.0<\eta<4.5$ and \N candidates with $\pt > 1 \gev$ and $\ptot > 15 \gev$. This selection achieves an efficiency exceeding $90\%$ for $m_{\N} \gtrsim 2 \gev$.

A multivariate analysis is performed to suppress combinatorial background, which dominates in the whole \muon\pion mass range, except for the region below 2\gev, where peaking backgrounds from charm meson decays are also present.
A neural network (NN) classifier, implemented using \textsc{PyTorch}~\cite{paszke2019pytorch}, is trained for each \textit{reconstruction category} in the analysis. These reconstruction categories are defined by each of the two muon charge configurations (SS and OS), three $\B$-decay modes (\decay{\B}{\muon \N \PX}, \decay{\Bp}{\muon \N}, and \decay{\Bc}{\muon \N}), and two analysis displacement categories (long and downstream). This gives nine reconstruction categories in total (no OS downstream data are collected). Each classifier is trained using simulated HNL samples as signal proxy and candidates in the upper \(m(\muon\muon\pion)\) sideband in data as background proxy. The fully reconstructed leptonic \decay{\Bp}{\muon \N}, and \decay{\Bc}{\muon \N} categories are combined for the purpose of the NN training.

For fully reconstructed leptonic decays,  the upper sideband corresponds to the \(m(\muon\muon\pion)\) region at least 3$\sigma$ (55.5\mev) above the \Bp mass peak, excluding the \Bc signal region defined as $\pm3\sigma$  ($\pm76.8$\mev) around the \Bc peak, with $\sigma$ the mass resolution.
Partially reconstructed \decay{\B}{\muon \N \PX} semileptonic decays are dominated by modes involving an unreconstructed $\D$ meson; therefore,  the sideband is defined as the region where $m(\muon\muon\pion) > m_\Bp-m_{\Dz}$.
A small fraction of signal \decay{\B}{\muon \N \PX} decays, with $\PX = \pion,\Prho$, may enter this sideband, but their contribution is expected to be negligible. To avoid biases, a 15\% portion of both the signal and background proxy datasets is reserved from training, and used solely to evaluate the NN's classification performance.

Variables with high discrimination power are used to train the classifiers. These include the \B-candidate flight distance, 
the \pt of all final-state particles, the \chisqip of the \N candidate and its decay products, the $\N$-candidate decay vertex $\chi^2$, and the angle between the momentum and displacement vector from the origin vertex to the decay vertex for the $\N$ candidate. For fully reconstructed leptonic decays, the angle between the momentum and displacement vector for the \B-meson candidate is also used. The simulated signal samples for the SS and OS categories show no significant differences and are therefore combined for training. In contrast, background samples show different kinematic characteristics between SS and OS categories and are treated separately.

To maximise discrimination power across the full mass range, particularly between simulated mass values, a novel NN-based approach is employed, in which the classifier is parametrised by a physics variable, the HNL mass~\cite{Baldi:2016fzo}.
In the NN training, the parametrised mass feature is set to the true value of the generated HNL mass for the signal proxy, while, for the background, it is arbitrarily sampled from the simulated mass values. This approach allows the classifier to smoothly interpolate between simulated mass values and eliminates the need for separate classifiers trained at individual masses.
The NN hyperparameters are tuned via Bayesian optimisation employing a Gaussian process~\cite{balandat2020botorch}, using classification accuracy as the optimisation metric. An independent set of optimal hyperparameter values is determined for each category. No sign of overtraining is observed in the classifier outputs.

Since the level of combinatorial background changes with the probed $m_{\N}$ values, multiple working points are defined to optimise the classifier's performance. For each simulated $m_{\N}$ value, the optimal requirement on the classifier output is determined and interpolated across the mass range. The working point is selected to maximise the Punzi figure of merit~\cite{Punzi:2003bu}, defined as
\begin{equation}
\mathrm{FOM}(m_{\N}) = \frac{\varepsilon(m_{\N})}{\sqrt{N_{\rm bkg}(m_{\N})} + \frac{3}{2}},
\end{equation}
where $\varepsilon(m_{\N})$ is the signal efficiency for a given $m_{\N}$ value.  The expected background yield, $N_{\rm bkg}(m_{\N})$, is extrapolated from the \(m(\muon\muon\pion)\) upper mass sideband into a window of $m_{\N} \pm 3\sigma_{m_{\N}}$, with $\sigma_{m_{\N}} = 25\mev$, corresponding to the largest mass resolution measured in fits across the simulated HNL mass values. Variations of $\sigma_{m_{\N}}$ yield consistent results. The optimal requirements are then relaxed, if necessary, to ensure that at least approximately 30 candidates are selected in the signal region, to ensure stable fits, even if this leads to a suboptimal FOM.

\section{Signal description}
\label{sec:sig_fit}

The signal yield $N_{\rm sig}$ expected in the selected data sample for each of the considered $\decay{\B_\mathrm{q}}{\muon \N (\PX)}$ decays, where $\rm q$ indicates the non-\bquark quark flavour, is given by
\begin{equation}
N_{\rm sig} = \mathcal{L}\,\sigma(\decay{\proton\proton}{\Bpm})\,\frac{f_\mathrm{q}}{f_\uquark}\,\mathcal{B}(\decay{\PB_\mathrm{q}}{\muon \N (\PX)})\,\mathcal{B}(\decay{\N}{\muon\pion}) \frac{\varepsilon(\FVN)}{\varepsilon(\FVBpm)}\, \varepsilon_{\rm sel},
\end{equation}
where the various factors are explained in the following.

The total integrated luminosity of the analysed data samples is ${\cal L} = 5.04\pm 0.10\invfb$. The inclusive cross-section at \sqs=13\tev, $\sigma(\decay{\proton\proton}{\Bpm}) = 86.6 \pm 6.4 \mub$, was measured by the LHCb experiment~\cite{LHCb-PAPER-2017-037} in the fiducial volume $\FVBpm$ defined by $2.0 < y(\Bpm) < 4.5$ and $ \pt(\Bpm) < 40 \gev$, where $y$ is the rapidity. 
The ratio of geometrical acceptances $\varepsilon(\FVN)/\varepsilon(\FVBpm)$ is introduced to correct for  the small mismatch between the fiducial volume selections.
Imposing the $\Bpm$ fiducial requirements on top of the analysis selection is found to be $>99.5\%$ efficient, indicating that any residual effect is negligible. The ratio is determined using the {\tt RapidSim} fast-simulation toolkit~\cite{Cowan:2016tnm}. For this, \B mesons are generated with a momentum spectrum predicted by fixed-order next-to-leading-log (FONLL) calculations~\cite{Cacciari:1998it}, and {\tt RapidSim} subsequently models the detector's geometric acceptance. This fast-simulation approach makes it feasible to determine the acceptance in fine steps of $m_{\N}$, which is crucial for modelling the sharp acceptance drops near phase-space boundaries. As only the ratio is used, common assumptions largely cancel, reducing uncertainties in the modelled B-meson production spectrum. The statistical uncertainty on the acceptance ratio is approximately 0.5\%, and the results are validated against the full \lhcb simulation at several benchmark $m_{\N}$ values.

The factor $f_\mathrm{q}/f_\uquark$ denotes the ratio of fragmentation fractions of the \bquark quark to the $\B_\mathrm{q}$ and $\Bpm$ mesons.
The ratio $f_\dquark/f_\uquark$ is assumed to be unity due to isospin symmetry, while $f_\squark/f_\uquark = 0.244 \pm 0.012$ and $f_\cquark/f_\uquark = (7.56 \pm 1.78)\times 10^{-3}$ are taken from LHCb measurements~\cite{LHCb-PAPER-2018-050, LHCb-PAPER-2019-033}. 
The $\Bs$ and $\Bc$ fragmentation fractions are determined in the range $4<\pt<25\gev$ and in the ranges $2.0<\eta<5.0$ and $2.5<\eta<4.5$, respectively. These  fiducial volumes are then slightly different to the $\Bpm$ cross-section fiducial volume. 
The dependence of the fragmentation functions on \pt and $\eta$ is small~\cite{LHCb-PAPER-2018-050, LHCb-PAPER-2019-033}, and thus any residual bias due to the fiducial volume difference is subleading compared to the quoted experimental uncertainties on $f_\mathrm{q}/f_\uquark$. As a result, the fiducial volume difference is neglected.

The signal selection efficiency $\varepsilon_{\rm sel}$ encodes the effects of all online and offline requirements for signal decays produced within the FV of the analysis and is determined in simulation as a function of HNL mass and lifetime. 
Efficiencies for alternative lifetimes are obtained from simulated samples through a weighting procedure, in which each candidate $i$ is assigned a weight 
\begin{equation}
\label{eq:tau_weight_simple}
    w^i_{\tau^\prime} \equiv \frac{\tau_{\rm gen}}{\tau^\prime} 
    \frac{e^{- t_i/\tau^\prime}}{e^{- t_i/\tau_{\rm gen}}}
\end{equation}
based on the generated lifetime of the sample, 
$\tau_{\rm gen}$, the target lifetime, $\tau^\prime$ and the true decay time of the candidate in the event, $t_i$.

To ensure the realistic modelling of the detector and trigger responses, several corrections are  applied. The simulation of the hardware trigger is corrected to match the performance observed in data. This correction is determined using a control sample of events containing muon pairs that were collected using triggers independent of those muons, ensuring an unbiased dataset for the efficiency measurement. The hardware trigger efficiency, $\varepsilon_{\rm L0}$, is then measured on this control sample by calculating the fraction of events in which the muons under study also satisfy the hardware trigger requirements. By comparing the efficiencies measured in data and simulation, correction weights, $w_{\rm L0}$, are derived. These weights are parametrised as a function of the maximum and minimum $\pt$ of the two signal muons to accurately adjust the simulation.

The efficiency of the GEC on the number of SPD hits, $n_{\rm SPD}$, is determined from simulation that is corrected to match the distribution in data. Specifically, the simulated SPD multiplicity is rescaled event-by-event by a global factor, $k_{n_{\rm SPD}}$, chosen such that the efficiency of the $n_{\rm SPD}<450$ requirement matches that observed in data measured with dimuon triggers using $n_{\rm SPD}<900$. After this correction, a 10\% relative uncertainty is assigned to the GEC efficiency to cover residual data-simulation differences; this is treated as a correlated nuisance across categories.

To correct for differences in the PID response between data and simulation, per-particle calibration weights, denoted as $w_{\text{PID}}$, are applied to the final-state particles of each simulated signal candidate. These weights are derived from high-statistics control samples~\cite{LHCb-PUB-2016-021} and are defined as the ratio of the PID efficiency measured in data to that in simulation. The weights are determined as a function of each particle's momentum and pseudorapidity to account for kinematic variations in the detector performance. The muon identification weights are calibrated using a pure sample of $\decay{\jpsi}{\mup\mun}$ decays, while the pion weights are derived using pions from \decay{\Dstarp}{\Dz(\to \Km\pip)\pip} decays selected only applying kinematic requirements. The total PID correction for a given signal candidate, $w_\text{PID}^i$, is the product of the individual weights applied to each of its three final-state particles.

The total selection efficiency $\varepsilon_{\rm sel}$ for a sample of a given HNL mass and lifetime is therefore defined as
\begin{equation}
    \varepsilon_{\rm sel}(m_\N, \tau^\prime) = \dfrac{\sum_i^{N_{\rm sel}} w^i_{\tau^\prime} w_{\rm L0}^i w_\text{PID}^i}{N_{\rm gen}} ,
\end{equation}
where $N_{\rm gen}$ and $N_{\rm sel}$ are the number of generated and selected signal events within the FV, respectively.
A cubic spline is employed to interpolate $\varepsilon_{\rm sel}$ between simulated HNL mass points. The selection efficiencies are determined for all reconstruction categories scanning the 2D plane of relevant HNL masses and lifetimes, in the ranges $m_{\N}\in[1.6,5.5]\gev$ and $\tau_\N\in[1\ps,10\ns]$, respectively.

For the long categories, the selection efficiencies $\varepsilon_{\rm sel}$ are maximal for values of $m_{\N}$ between 3 and $4 \gev$
where the HNL is heavy enough to provide high transverse momentum to the secondary muon, yet light enough to allow the primary muon (from the \B meson) to also have a relatively large \pt.
Moreover, the efficiencies are significantly larger for \decay{\Bp}{\mup \N} than for \decay{\Bc}{\mup \N} because the \Bp lifetime is about three times larger than the \Bc lifetime, making the impact parameter requirements more effective. 
On the other hand, the efficiency for the downstream category is larger for low $m_{\N}$ as the primary muon gets a larger share of the energy, and the first-stage software trigger is more efficient.

\begin{figure}[tb]
    \centering
    \includegraphics[width=.49\textwidth]{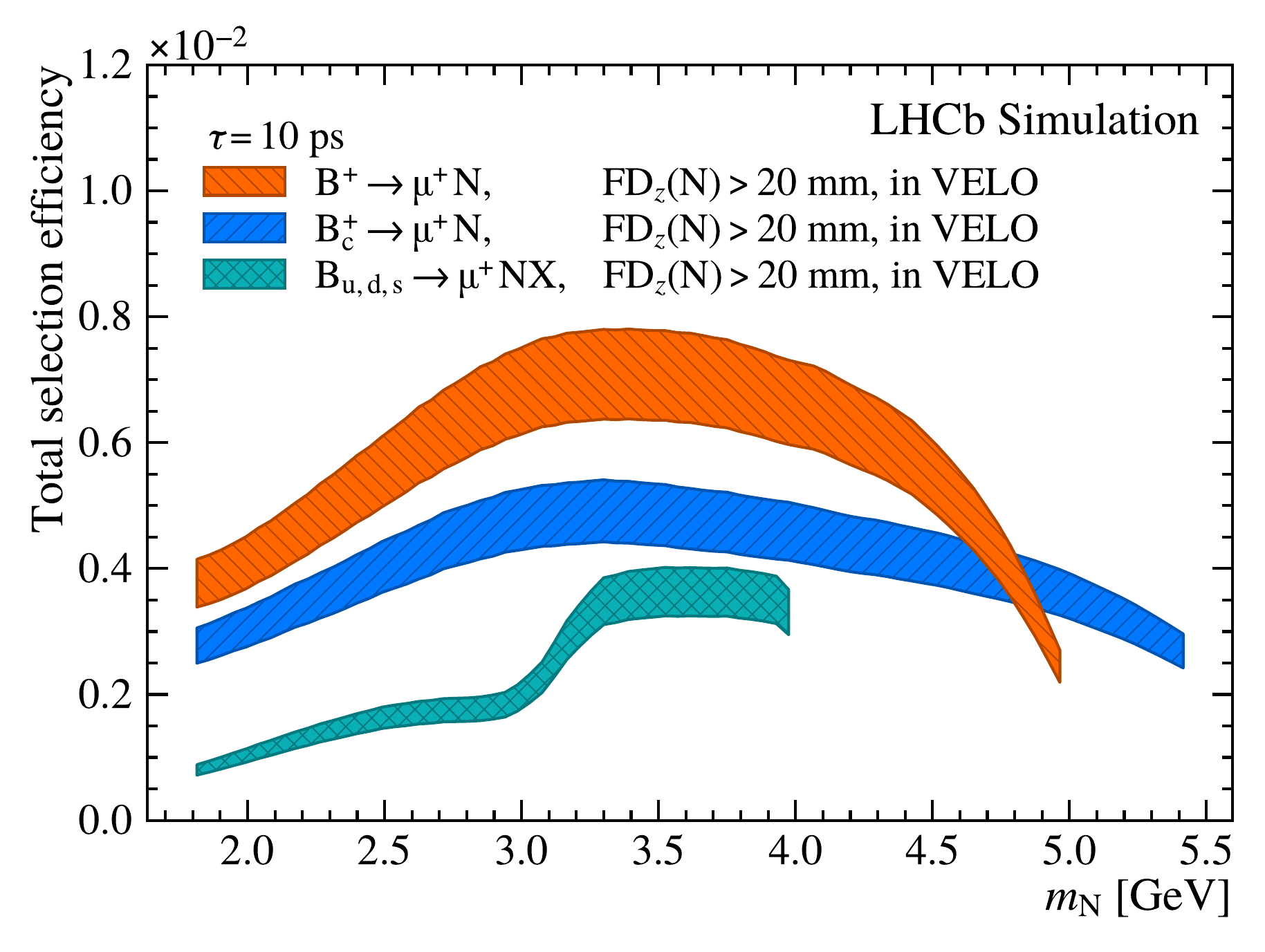}
    \includegraphics[width=.49\textwidth]{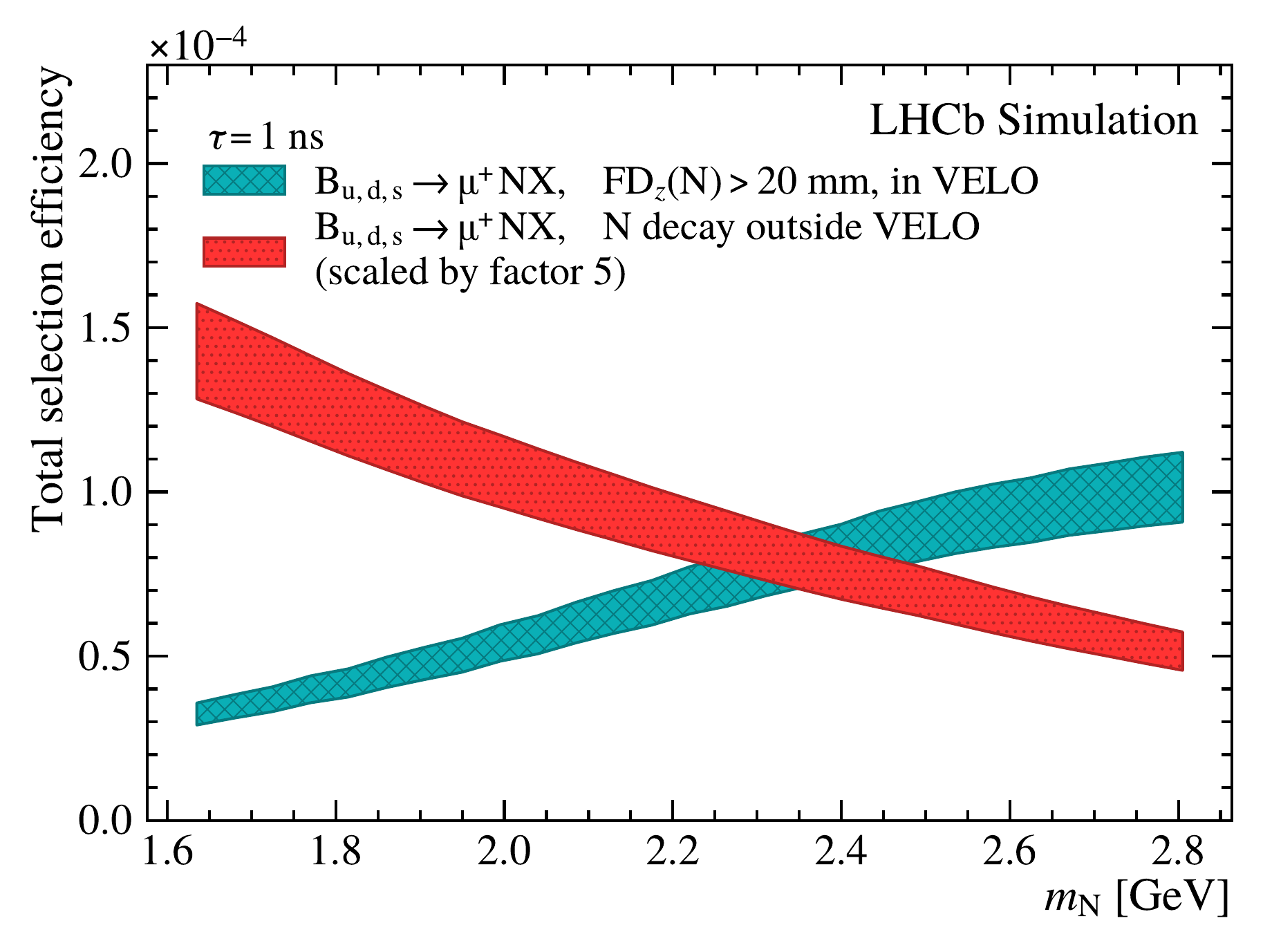}
    \caption{
    Total selection efficiency for HNL events with two same-sign muons, shown for HNL lifetimes of (left) 10\ps  and (right) 1\ns. Coloured bands represent the allowed range of values within $\pm1\sigma$ of the combined statistical and systematic uncertainties. For a direct comparison between reconstruction categories, the downstream-track efficiency is multiplied by a factor of five, corresponding to the rescaling factor of the dedicated trigger.} 
    \label{fig:SelectionEfficiency}
\end{figure}

As an example, the total selection efficiencies for the different same-sign reconstruction modes are shown in Fig.~\ref{fig:SelectionEfficiency} for HNL lifetimes of 10\ps and 1\ns as a function of $m_\N$. Opposite-sign efficiencies are similar. For a lifetime of 10\ps, HNLs are sufficiently short-lived that downstream reconstruction is not efficient. At a lifetime of 1\ns, the analysis is only sensitive to the more abundantly-produced light ($m_{\N} < 3\gev$) HNLs  due to a weaker HNL-SM coupling. In this regime, HNL production is dominated by $\decay{\B_{\uquark,\dquark,\squark} }{ \mup \N \PX}$ decays. 
As a result, and for reasons of legibility, the downstream category is not shown on Fig.~\ref{fig:SelectionEfficiency} (left), and, similarly, the decays $\Bcp \rightarrow \mup \N$ and $\B^+ \rightarrow \mup \N$ are not shown on Fig.~\ref{fig:SelectionEfficiency} (right). 
An efficiency of the order of 0.5\% is achieved for the 10\ps lifetime, with comparable values for all production modes of the long category. For the 1\ns lifetime, the total efficiency is highly reduced. 
After correcting for the prescale factor of five, and at low $m_\N$, the downstream-track selection is approximately three times as efficient as the long-track selection for an HNL lifetime of 1\ns.
Due to the relatively large \decay{\bquark}{\cquark} coupling, most \decay{\B_{\uquark,\dquark,\squark}}{\mup\N{\rm X}} decays will include a charm meson in the \B decay products when kinematically allowed. This produces a softer momentum spectrum for HNL masses below this kinematic threshold, and a harder spectrum for those above, where the accompanying decay products are much lighter. This is the cause of the kink in the \decay{\B_{\uquark,\dquark,\squark}}{\mup\N{\rm X}} efficiency at around 3\gev.

\begin{figure}[tb]
    \centering
    \includegraphics[width=.49\textwidth]{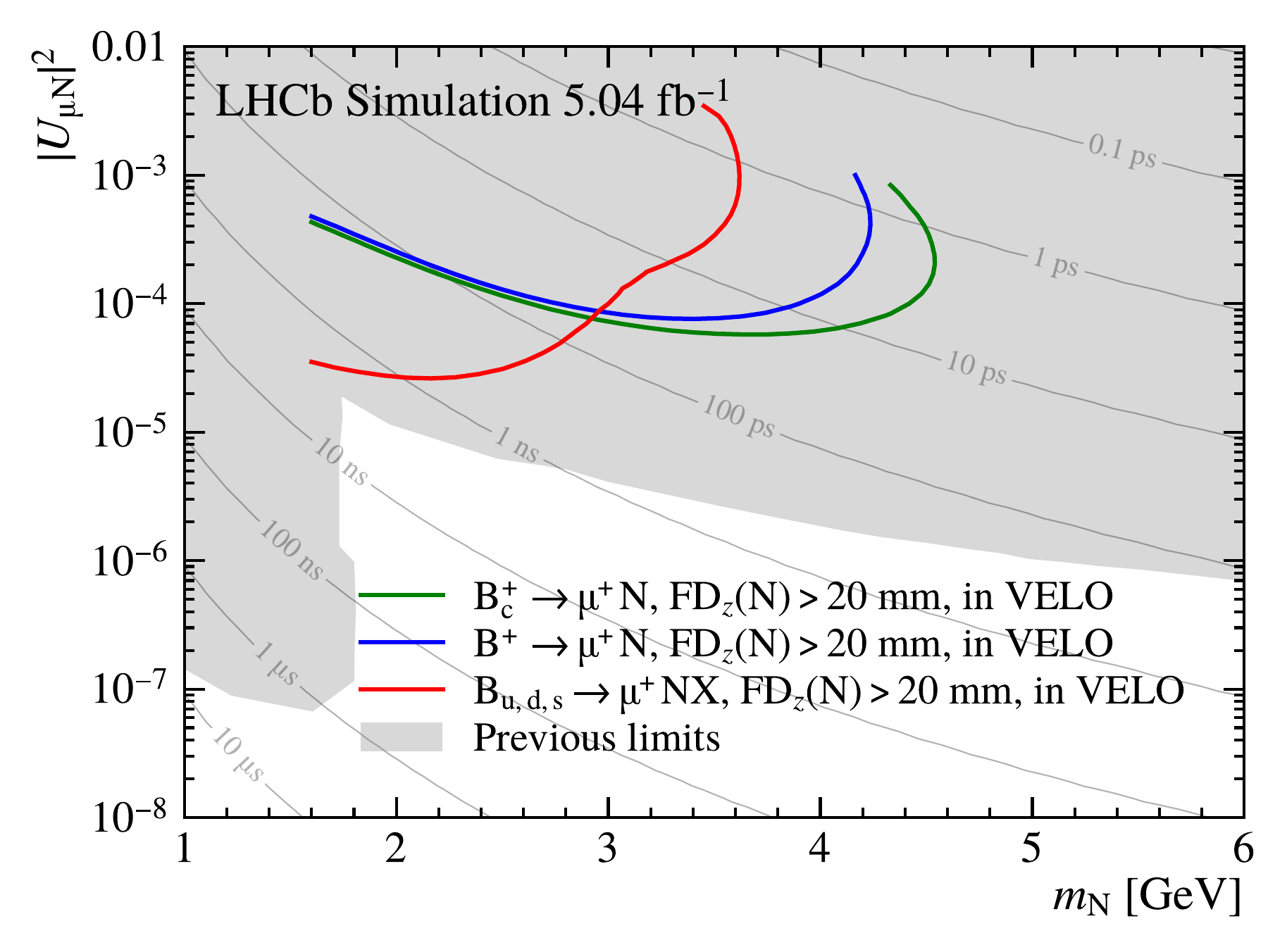}
    \includegraphics[width=.49\textwidth]{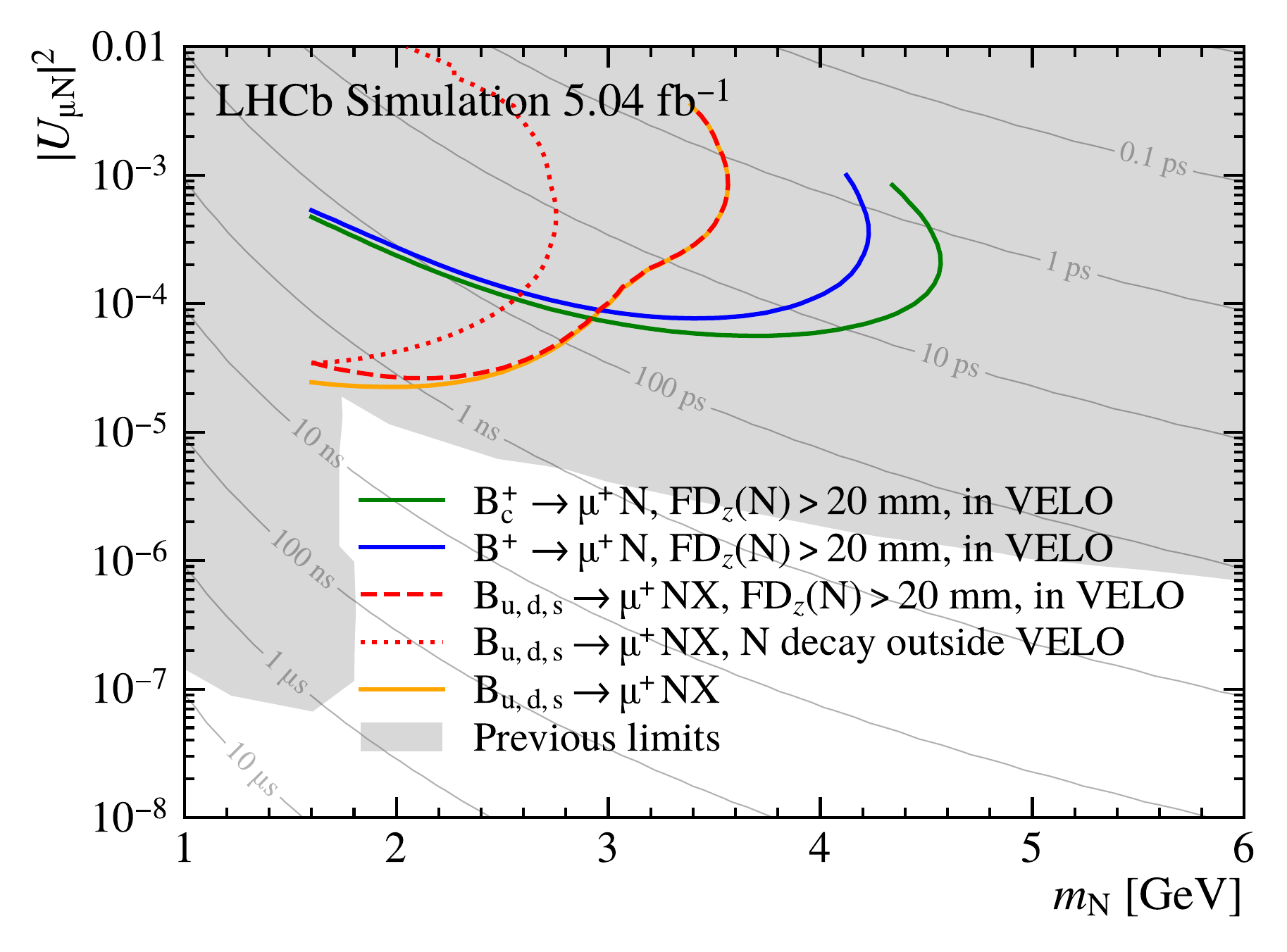}
    \caption{Sensitivity estimate of various reconstruction categories of the analysis for the (left) OS and (right) SS event categories, assuming no background. Lifetimes, shown by the grey contours, and previous limits are for a Majorana-natured HNL. The most stringent limits in this mass range are from the CMS and BEBC collaborations.}  
    \label{fig:zero_bkg}
\end{figure}

Figure~\ref{fig:zero_bkg} shows the  sensitivity of each reconstruction category, obtained assuming no background. Below 3\gev, the semileptonic \textit{long} categories are expected to have the most competitive sensitivity in both SS and OS signatures. The contribution of the \textit{downstream} category is significant for HNL masses of about 2\gev. The sensitivity from fully reconstructed leptonic categories becomes dominant above 3\gev.

\section{Background estimation}
\label{sec:bkg_fit}

The combinatorial background is described by a simple smooth function (exponential or linear), whose parameters are determined in the final fit in a sliding $\muon\pion$ mass window, as described in Sec.~\ref{sec:sig_ext}. 
A peaking background contribution due to \decay{\Dz}{\Km\pip} and \decay{\Dz}{\pim\pip} decays is estimated for the semileptonic \textit{long} category. In these \D decays, a peak appears in the \muon\pion invariant mass through the misidentification of the \Km, \pim or \pip as a muon. In fully reconstructed leptonic decays, more stringent requirements on the \muon\muon\pion invariant mass suppress the misidentified background. The peaking background contribution is found to be negligible for the semileptonic \textit{downstream} category due to the short lifetime of charm mesons. 
A data-driven \texttt{ABCD} method~\cite{Behnke:2013pga,D0:1994wmk,PhysRevD.44.29} (named after the four subsamples used in this approach, labelled A, B, C, and D) is employed to estimate these charm peaking background yields. The background levels from a control region of prompt-like decays, \ie candidates with $\mathrm{FD}_z<20$\mm, are extrapolated into the signal region.
The use of this prompt control region is justified because the parameter space for prompt HNLs with masses up to 2\gev is already strongly constrained by dedicated searches at other experiments~\cite{Liventsev:2013zz,Belle:2022tfo,CMS:2018iaf,Aad:2019kiz}. Therefore, any prompt candidates in this mass range are expected to originate predominantly from charm meson decays rather than genuine HNL signals, making this region suitable as a background-dominated control sample for the \texttt{ABCD} method.

\begin{figure}[tb]
    \centering
    \includegraphics[width=.49\textwidth]{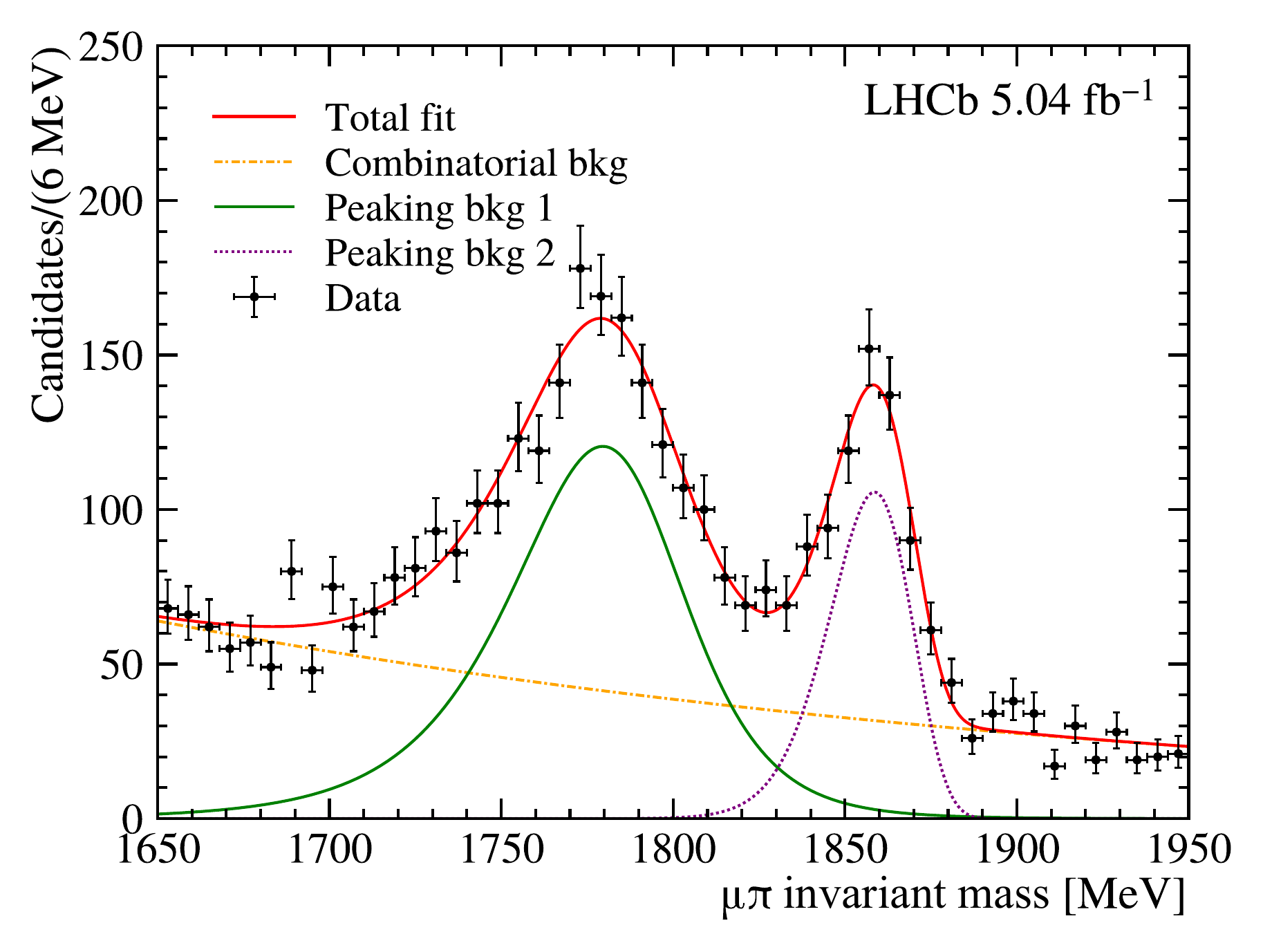}
    \includegraphics[width=.49\textwidth]{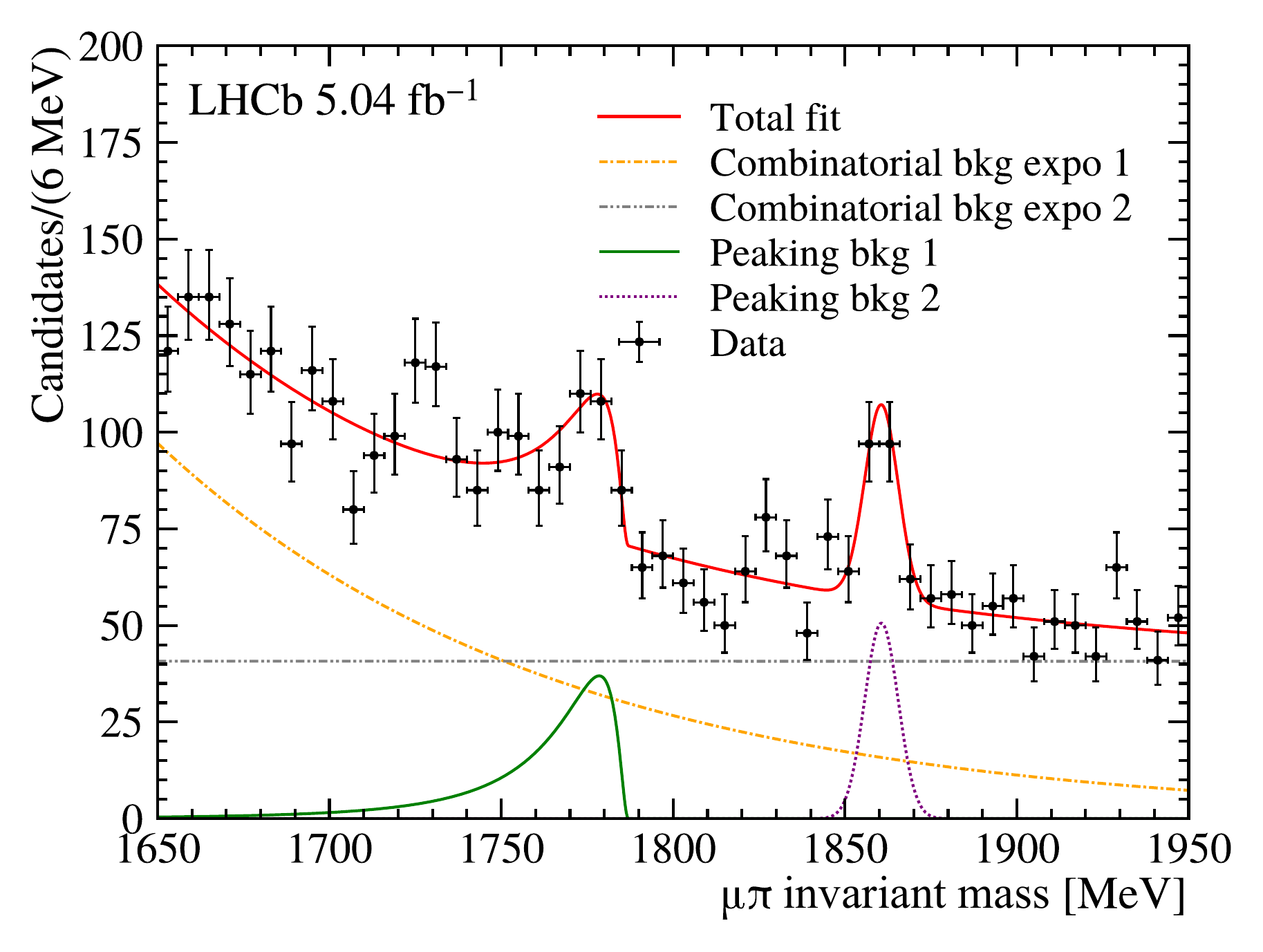}
    \caption{
    Distribution of the \(\muon\pion\) mass in the control region D for the (left) SS and (right) OS \textit{long} categories, showing fits with two Johnson \(S_{\mathrm{U}}\) distributions for peaking backgrounds from charm meson decays and  either one (left) or two (right) exponentials for the combinatorial background.} 
    \label{fig:pkbkg_fit}
\end{figure}

The signal and control regions are defined using two statistically independent variables: the HNL flight distance $ \mathrm{FD}_z(\N) $ and the invariant mass $ m(\muon\muon\pion) $. The $ \mathrm{FD}_z(\N) $ variable divides the sample into displaced (regions A and C, with $ \mathrm{FD}_z(\N) \geq 20 $ mm) and prompt (regions B and D, with $ \mathrm{FD}_z(\N) < 20 $ mm) categories, while the invariant mass $ m(\muon\muon\pion) $ divides the sample into signal (regions A and B) and sideband (regions C and D) regions. Mass peaks from $ \decay{\Dz}{\Km\pip} $ and $ \decay{\Dz}{\pim\pip} $ are observed in the control regions. The two peaks in the $ \muon\pion $ spectrum in these categories are described with two Johnson $ S_{\mathrm{U}} $ distributions~\cite{JohnsonSU}, while the combinatorial background is modelled with an exponential function (two for the OS case). Figure~\ref{fig:pkbkg_fit} shows the data distributions in region D for the SS and OS categories, with the corresponding fits.

Peaking-background yields in region A are predicted with an \texttt{ABCD} extrapolation, $N_{\rm A}^{\text{pred}} = N_{\rm B} \times (N_{\rm C}/N_{\rm D})$, where $N_X$ are the peaking yields obtained from invariant-mass fits in regions $X$. Their full uncertainties are propagated. As a result, $N_{\rm A}^{\text{pred}}$ is found to be consistent with zero, but is nonetheless used as a constrained nuisance parameter in the final mass fit. The shape parameters of the Johnson $ S_{\mathrm{U}} $ probability density function (PDF) are measured in the most populated region D and fixed when fitting regions A, B, and C. The exponential shape parameters are free to vary in each region. The statistical independence of $ \mathrm{FD}_z(\N) $ and $ m(\muon\muon\pion) $ is verified by applying the \texttt{ABCD} method to a subdivision of regions B and D.

\section{Signal extraction}
\label{sec:sig_ext}

To search for evidence of HNLs, a bump hunt is performed blindly in sliding windows across the whole $m(\muon\pion)$ spectrum. The HNL mass hypotheses in this spectrum, $m_{\N}$, are scanned in steps of $\frac{1}{2} \sigma_{m_{\N}}$, where $\sigma_{m_{\N}}$ is the mass resolution at each $m_{\N}$ value. For each hypothesis, the local search window is defined as  $|m - m_{\N}| < 7 \sigma_{m_{\N}}$.

The size of the local search window, the signal mass model, and the mass steps depend on the value of $m_{\N}$. For a set of mass points, the signal mass model is extracted from simulated samples and modelled by a modified Crystal Ball function~\cite{Skwarnicki:1986xj} with tails on both sides of the peak. Between simulated mass points, the signal PDF is interpolated using the moment-morphing technique~\cite{Baak:2015}. No dependence of the signal shape on lifetime is observed.

The combinatorial background is modelled using an exponential function, $\mathcal{B}_{\rm comb}(m, N_{\rm comb}, \alpha_{\rm comb}) \propto e^{- \alpha_{\rm comb} m}$, which is treated as the \textit{baseline} model. For a small decay constant, this function approximates a linear shape. To assess the systematic uncertainty associated with this choice, an \textit{alternative}, explicitly linear, model, $\mathcal{B}_{\rm comb}(m,N_{\rm comb}, \alpha_{\rm comb}) \propto (1 + \alpha_{\rm comb} m)$, is used. The difference between the results obtained with the baseline and alternative models is used to quantify the corresponding systematic, as detailed in Sec.~\ref{sec:syst}. In both cases, the normalisation $N_{\rm comb}$ and the shape parameter $\alpha_{\rm comb}$ are allowed to vary in each fit and are independent between the analysis categories.

When the search window overlaps with the peaking background region, defined between 1650 and 1950\mev, two peaking background PDFs are also included for the semileptonic \textit{long} categories of the search (both SS and OS). Their shape parameters are fixed as described in the previous section, while the expected yield from the \texttt{ABCD} method is incorporated as a Gaussian prior.

For each interpretation of the HNL nature (Dirac or Majorana), the different reconstruction categories, $i$, are fitted simultaneously using a common yield parameter, $N_{\rm sig}$. The total PDF for each reconstruction category is expressed as
\begin{equation}
\label{eq:s_and_b}
\mathcal{P}_i(m; m_{\N}, N_{\rm sig}, \theta ) = \frac{N_{\rm sig} \epsilon_i \mathcal{P}^{\rm sig}_i(m; m_\N, \theta_{\rm sig}) + \sum_{k=1}^3N_{{{\rm bkg},k}:i}\mathcal{P}^{{\rm bkg},k}_{i}(m; m_{\N}, \theta_{\rm bkg})}{N_{\rm sig} \epsilon_i+\sum_{k=1}^3N_{{{\rm bkg},k}:i}},
\end{equation}
with $k$ distinguishing the three components of background: combinatorial ($k=1$), and the two peaking backgrounds ($k=2,3$). In this equation, $\mathcal{P}^{\rm sig}_i$ and the set of $\mathcal{P}^{{\rm bkg},k}_{i}$ are the signal and background model PDFs, $\epsilon_i$ is the fraction of all HNL candidates that are in reconstruction category $i$,  the set of $N_{{\rm bkg},k:i}$ are the background yields in category $i$ and background component $k$, and the nuisance parameters (exponential slopes, \etc) are denoted by $\theta \equiv(\theta_{\rm sig}, \theta_{\rm bkg})$. When peaking background is not present, \ie for all reconstruction categories except the semileptonic long categories, or when the peaking background is outside of the HNL mass window, the peaking background yield is forced to be identically zero, \ie $N_{{\rm bkg},2:i}=N_{{\rm bkg},3:i}=0$.

The extended likelihood is  
\begin{equation}
\label{eq:bin_likelihood}
    \mathcal{L}_i(m_{\N},N_{\rm sig}, \theta) = \frac{N_i^{N_{{\rm obs}:i}} e^{-N_i}}{N_{{\rm obs}:i}!} \mathcal{C}_{\epsilon_i}(\theta_{\rm sig}) \mathcal{C}_{{\rm bkg}:i}(\theta_{\rm bkg}) \prod_{j=1}^{N_{{\rm obs}:i}}\mathcal{P}_i(m_j; m_{\N}, N_{\rm sig}, \theta )\,,
\end{equation}
where $N_{{\rm obs}:i}$ is the observed number of candidates in the reconstruction category,  $N_i$ is the total yield; $N_i  = N_{{\rm sig}:i} + \sum_{k=1}^3N_{{\rm bkg},k:i}$, and $m_j$ the invariant mass of observed event $j$. The $\mathcal{C}_{\epsilon_i}$ factor represents the Gaussian prior on the event fraction in reconstruction category $i$
\begin{equation}
\label{eq:effcy_constraint}
    \mathcal{C}_{\epsilon_i}(\theta_{\rm sig})  = \frac{1}{\sqrt{2\pi} \sigma_{\epsilon_i} } \exp \left(-\frac{(\mu_{\epsilon_i} - \epsilon_i)^2}{2\sigma_{\epsilon_i}^2} \right),
\end{equation}
where the central value, $\mu_{\epsilon_i}$, and uncertainty, $\sigma_{\epsilon_i}$, of $\epsilon_i$ are obtained from simulation. The factors $\mathcal{C}_{{\rm bkg}:i}$ are the Gaussian priors on the peaking backgrounds normalisations, defined as

\begin{equation}
\label{eq:peak_bkg_constraint}
    \mathcal{C}_{{\rm bkg}:i}(\theta_{\rm bkg})  = \prod_{k=2}^3\frac{1}{\sqrt{2\pi} \sigma_{N_{{\rm bkg},k:i}} } \exp \left(-\frac{(\mu_{N_{{\rm bkg},k:i}} - N_{{{\rm bkg},k:i}})^2}{2\sigma_{N_{{\rm bkg},k:i}}^2} \right),
\end{equation}
when reconstruction categories contain peaking background, and set to one elsewhere. In this definition, $\mu_{N_{{\rm bkg},k:i}}$ and $\sigma_{{N_{{\rm bkg},k:i}}}$ are the central values and uncertainties of the peaking background yield. 

The combined likelihood for the simultaneous fit of  different reconstruction categories is expressed by
\begin{equation}
\label{eq:likelihood_combined}
\mathcal{L}(m_{\N},N_{\rm sig}, \theta) = \prod_{i=1}^{N_{\rm cat}} \mathcal{L}_i(m_{\N},N_{\rm sig}, \theta).
\end{equation}

\begin{figure}[tb]
    \centering
    \includegraphics[width=.49\textwidth]{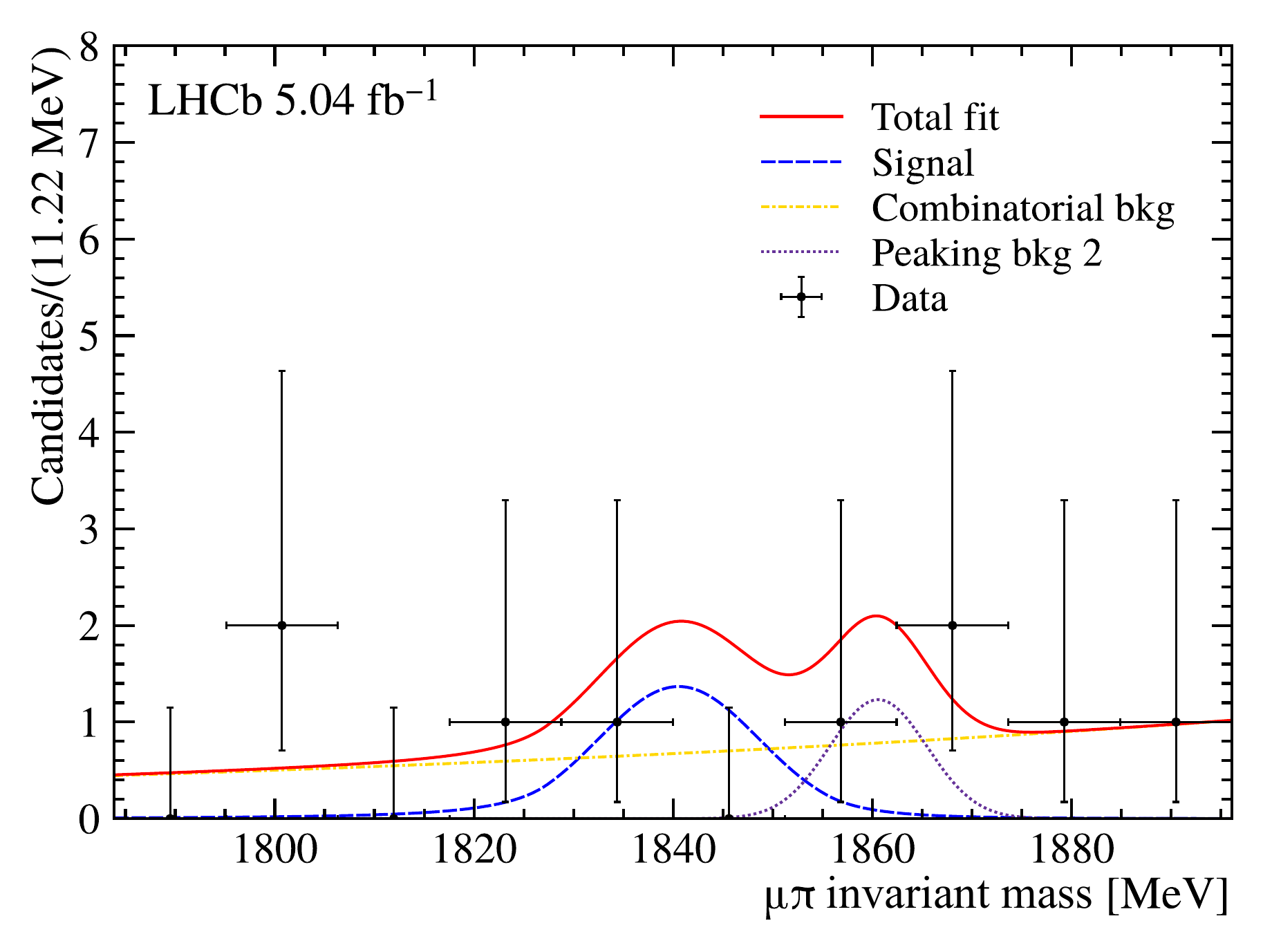}
    \includegraphics[width=.49\textwidth]{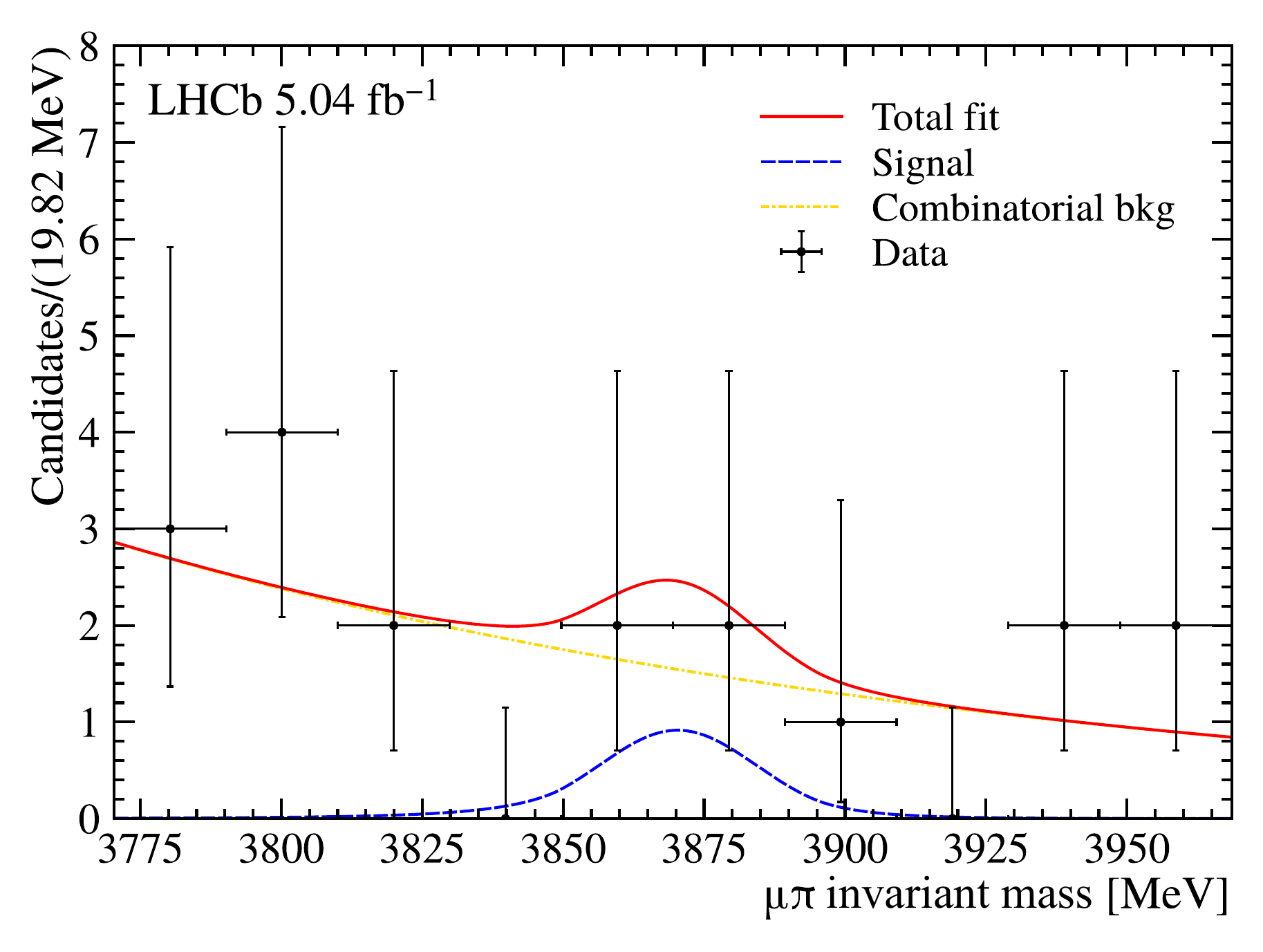}
    \caption{
    Examples of mass distributions obtained in the signal region for the OS \textit{long} category, with fits for Majorana HNLs. The plots correspond to the inclusive (left) and the leptonic \Bp (right) reconstruction categories.}  
    \label{fig:signal_fit}
\end{figure}

The fit is implemented using the \textsc{zfit} library \cite{Eschle:2020}, and yields are not constrained to be positive.
Examples of fits to the $\muon\pion$ mass spectra in the signal region are shown in Fig.~\ref{fig:signal_fit} for two  mass ranges: one with peaking background contributions (left) and one without (right).

For each mass hypothesis, several values of the $|U_{\muon \N}|^2$ coupling are tested to determine upper limits. These tested couplings, denoted $|U_{\muon \N}|^2({\rm tested})$, are used to compute the signal efficiency in each reconstruction category, which depends on the HNL lifetime and therefore on $|U_{\muon \N}|^2$.
The CL\textsubscript{s} method \cite{Cowan:2011} is used to translate the fitted signal yield $N_{\rm sig}$ into a 95\% confidence level (CL) upper limit on the muonic HNL coupling, $|U_{\muon \N}|^2$, evaluated at the fixed signal efficiency. This value is denoted as $|U_{\muon \N}|^2({\rm measured})$. The 95\% CL exclusion region is defined as the region where $|U_{\muon \N}|^2({\rm measured})$ lies below $|U_{\muon \N}|^2({\rm tested})$.
To determine the global significance of any deviation, the look-elsewhere effect is accounted for following the method of Ref.~\cite{Gross:2010qma}. In essence, the number of crossings of the observed limit curve over the expected limit is used to estimate the effective number of statistically independent test masses considered.

\section{Systematic uncertainties}
\label{sec:syst}

The systematic uncertainties are summarised in Table~\ref{tab:systematics_combined_final}. Relative uncertainties are quoted on the squared coupling, $|U_{\muon \N}|^2$, except for the fragmentation fractions, which are given relative to their baseline values (since their impact on $|U_{\muon \N}|^2$ is mass dependent). All systematic uncertainties enter the likelihood as Gaussian priors.

The leading external systematic uncertainties arise from the inclusive $\sigma(\decay{\proton\proton}{\Bpm})$ cross-section (7.4\%) and the $\Bc$ fragmentation fraction (24\% for HNLs from $\Bc$). The integrated luminosity (2.0\%) and the $\Bs$ fragmentation ratio (4.9\%) are also included as external constraints. The leading detector-related uncertainty is assigned to the GEC efficiency (10\%), derived from the data-driven $k_{n_{\rm SPD}}$-scaling procedure and the observed spread across HNL masses. 

The contribution due to the signal mass modelling is evaluated with pseudoexperiments by varying the signal-shape parameters according to their covariance from simulation fits; this yields a 5.9\% relative impact on $|U_{\muon \N}|^2$. The background parametrisation uncertainty is assessed by generating samples with one functional form and fitting with an alternative form, giving a contribution of 2.0\%.

Statistical uncertainties in the efficiency corrections (L0, PID) are propagated as additional Gaussian nuisances. External inputs and the GEC efficiency are taken as correlated across categories; shape-related nuisances are treated as uncorrelated between categories.

\begin{table}[tb]
\centering
\caption{
Summary of systematic uncertainties on the HNL coupling \(|U_{\muon \N}|^2\), including baseline values and relative uncertainties in percent. Relative uncertainties, when reported, are defined based on the impact on \(|U_{\muon \N}|^2\). Relative uncertainties for fragmentation functions are not reported as this depends on the HNL mass considered.}
\label{tab:systematics_combined_final}
\resizebox{1\columnwidth}{!}{
\begin{tabular}{lr@{\:$\pm$\:}lc}
\hline
{Source} & \multicolumn{2}{c}{Baseline value} & {Rel. uncertainty [\%]} \\
\hline
$\sigma(\proton\proton \to \Bp)$ ~$[\mub]$& 86.6 & 6.4 & 7.4 \\
Integrated luminosity $[\invfb]$ & 5.04 & 0.10 & 2.0 \\
$\Bcp$ fragmentation fraction ($f_\cquark/f_\uquark$) ~$[\times10^{-3}]$& 7.6  & 1.8 & \multicolumn{1}{c}{---} \\
\Bs fragmentation fraction ($f_\squark/f_\uquark$) & $0.244$ & $0.012$ & \multicolumn{1}{c}{---} \\
GEC efficiency & \multicolumn{2}{c}{---} & 10 \\
Signal shape modelling & \multicolumn{2}{c}{---} & 5.9 \\
Background shape modelling & \multicolumn{2}{c}{---} & 2.0 \\
\hline
\end{tabular}
}
\end{table}

\section{Results and conclusions}

\begin{figure}[tb]
    \centering
    \raisebox{-.3\height}{\includegraphics[width=.49\textwidth]{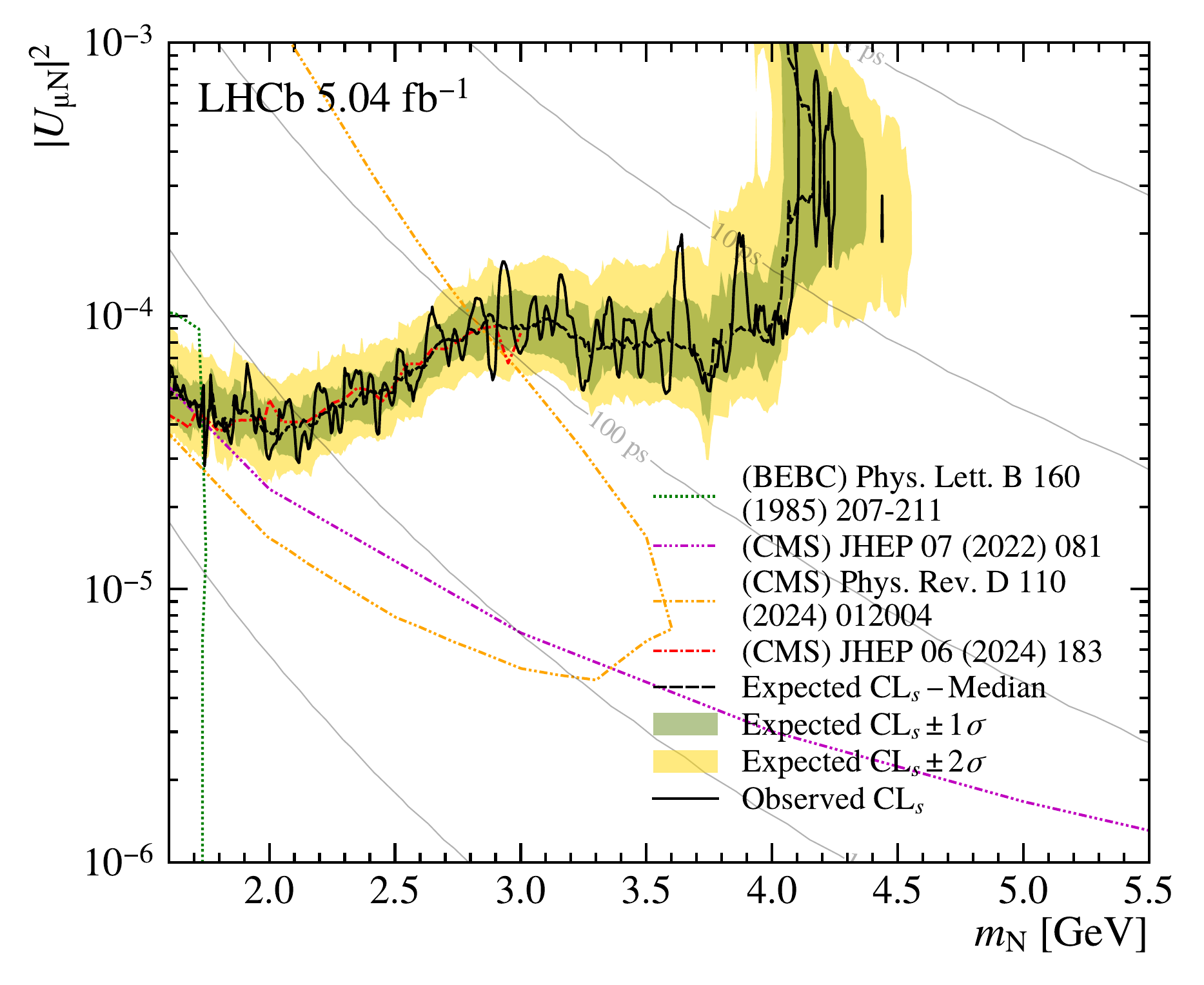}}
    \raisebox{-.3\height}{\includegraphics[width=.49\textwidth]{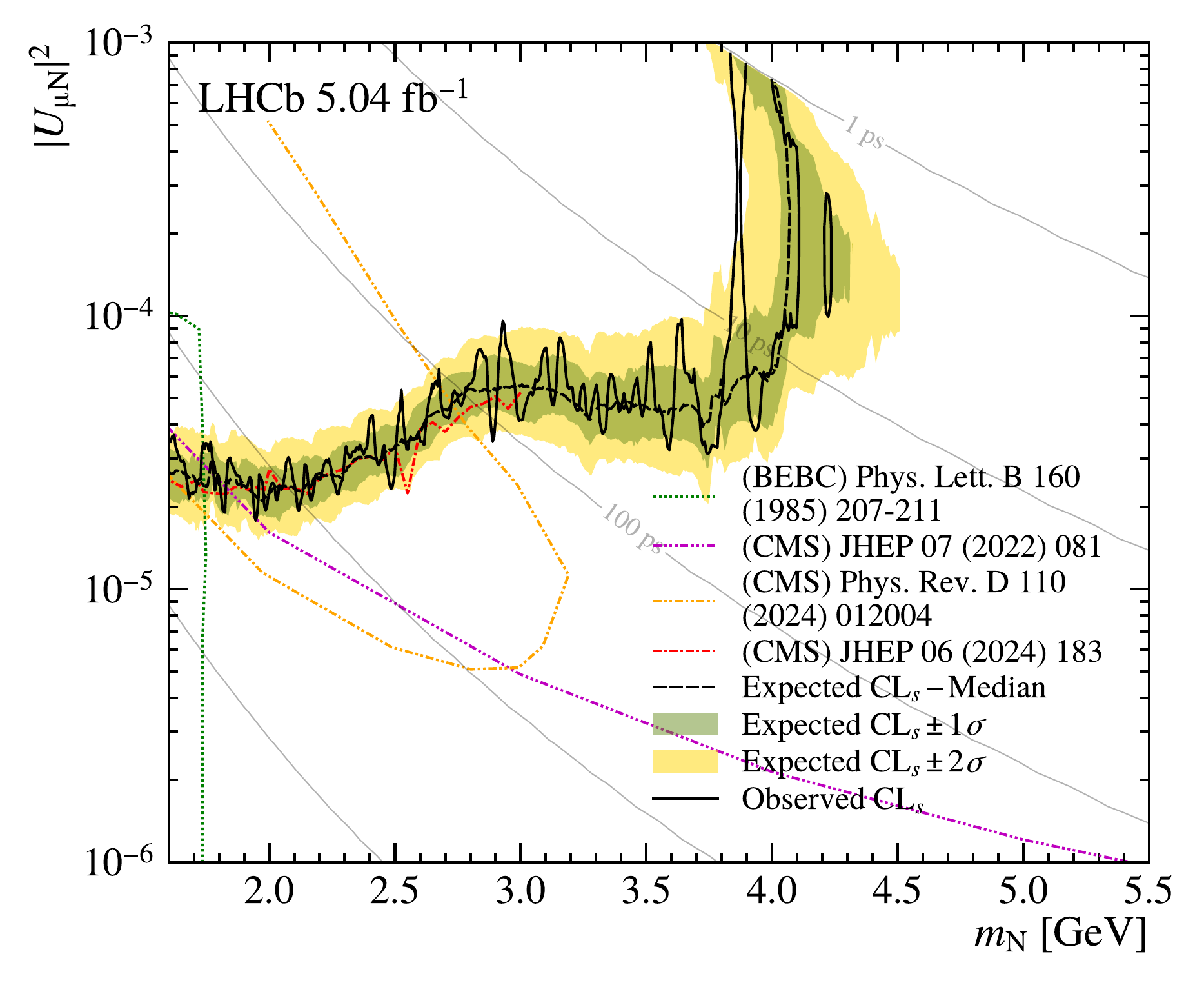}}
    \caption{(Left) Combination of all the OS search categories that corresponds to a Dirac-like HNL. (Right) Combination of all the SS and OS search categories that corresponds to a Majorana-like HNL. Lifetime contours are shown in grey.
}
    \label{fig:result_unblinded_30evt}
\end{figure}

A search for HNLs has been performed using a dataset corresponding to an integrated luminosity of ${\cal L} = 5.04\pm 0.10\invfb$ of proton-proton collision data collected with the \lhcb detector between 2016 and 2018. The search targets HNLs produced in beauty-meson decays with subsequent decays into a $\mupm\pimp$ final state, probing HNL masses in the range of 1.6 to 5.5\gev and their coupling to muon neutrinos.

The analysis combines fully reconstructed leptonic decays, $\decay{\B^+_{(c)}}{\mup \N}$, and partially reconstructed inclusive semileptonic decays, $\decay{\B}{\muon^+ \N X}$. Analysing final states with both same-sign and opposite-sign muons ensures sensitivity to both Majorana and Dirac HNL scenarios. A machine learning approach, featuring a neural network parametrised by the HNL mass, enhances signal discrimination. Using both \emph{long} and \emph{downstream} track categories provides sensitivity across a wide range of HNL lifetimes, corresponding to decay lengths from approximately 2\cm to 2\m.

The highest local excess has a $2.5\sigma$ significance and is observed at $m_{\N}=2.93\gev$ in the Dirac OS case. Taking into account the look-elsewhere effect, the probability of observing a local fluctuation of at least $2.5\sigma$ is estimated to be $38\%$, lowering it to a global statistical significance of $0.3\sigma$ (one-sided). Therefore, the observed results are compatible with statistical fluctuations.

Upper limits are set at the 95\% confidence level on the mixing parameter squared, $|U_{\muon \N}|^2$, as a function of the HNL mass $m_{\N}$, and are reported in Fig.~\ref{fig:result_unblinded_30evt}. These limits exclude values of $|U_{\muon \N}|^2$ down to the $10^{-5}$--$10^{-4}$ level. This represents more than an order of magnitude improvement over past LHCb limits. The limits are a valuable cross-check to the existing limits in this region. They also complement searches from ATLAS and CMS which are typically more sensitive at higher masses from on-shell $\W$ boson decays, and previous fixed-target experiments where all masses above the D-meson threshold ($\sim 1.7 \gev$) are inaccessible. 
The limits reported here do not probe the leptogenesis-favoured $\nu$MSM region but exclude significant parameter space in scenarios with three HNLs or additional interactions, such as symmetry-protected low-scale seesaws, flavour-structured models with dominant muon mixing, or frameworks with gauged $B-L$ (baryon minus lepton number), which predict larger mixings compatible with light-neutrino masses and cosmological constraints~\cite{Drewes:2021nqr,Drewes:2024bla}. 

The techniques developed, including peaking background control in inclusive \B decays, the treatment of downstream tracks, and a mass-parametrised neural network, provide a robust framework for future HNL searches. Significant sensitivity improvements~\cite{Alimena2025feeblyinteractingparticlesfips} are expected utilising advancements in  displaced vertex reconstruction~\cite{LHCb-DP-2022-001} and the larger datasets from the \lhcb Upgrade I~\cite{LHCb-DP-2022-002} and II~\cite{LHCb-TDR-026}.

% Do not include this in any draft (just for information in the template)
%\input{acknowledgements_intro}
% Comment this in for paper drafts; do not include this in analysis note, conference and figure reports
\section*{Acknowledgements}
%
% These Acknowledgements valid from 12/09/25
%
\noindent We express our gratitude to our colleagues in the CERN
accelerator departments for the excellent performance of the LHC. We
thank the technical and administrative staff at the LHCb
institutes.
We acknowledge support from CERN and from the national agencies:
ARC (Australia);
CAPES, CNPq, FAPERJ and FINEP (Brazil); 
MOST and NSFC (China); 
CNRS/IN2P3 (France); 
BMFTR, DFG and MPG (Germany);
INFN (Italy); 
NWO (Netherlands); 
MNiSW and NCN (Poland); 
MCID/IFA (Romania); 
%MSHE (Russia); 
MICIU and AEI (Spain);
SNSF and SER (Switzerland); 
NASU (Ukraine); 
STFC (United Kingdom); 
DOE NP and NSF (USA).
%%%%%%%%%%%%%%%%%%%%%%%%%%%%%%%%%%%%%%%%%%%%%
We acknowledge the computing resources that are provided by ARDC (Australia), 
CBPF (Brazil),
CERN, 
IHEP and LZU (China),
IN2P3 (France), 
KIT and DESY (Germany), 
INFN (Italy), 
SURF (Netherlands),
Polish WLCG (Poland),
IFIN-HH (Romania), 
%RRCKI and Yandex LLC (Russia), 
PIC (Spain), CSCS (Switzerland), 
and GridPP (United Kingdom).
%%%%%%%%%%%%%%%%%%%%%%%%%%%%%%%%%%%%%%%%%%
We are indebted to the communities behind the multiple open-source
software packages on which we depend.
%%%%%%%%%%%%%%%%%%%%%%%%%%%%%%%%%%%%%%%%%%
Individual groups or members have received support from
%ARC and ARDC (Australia); % moved to national 16/01
Key Research Program of Frontier Sciences of CAS, CAS PIFI, CAS CCEPP, 
%Fundamental Research Funds for the Central Universities,  and Sci.\
%\& Tech.\ Program of Guangzhou (China); Removed 24/11/25
Minciencias (Colombia);
EPLANET, Marie Sk\l{}odowska-Curie Actions, ERC and NextGenerationEU (European Union);
A*MIDEX, ANR, IPhU and Labex P2IO, and R\'{e}gion Auvergne-Rh\^{o}ne-Alpes (France);
%RFBR, RSF and Yandex LLC (Russia);
Alexander-von-Humboldt Foundation (Germany);
ICSC (Italy); 
%GVA, XuntaGal, GENCAT, Inditex, InTalent and Prog.~Atracci\'on Talento, CM (Spain);
Severo Ochoa and Mar\'ia de Maeztu Units of Excellence, GVA, XuntaGal, GENCAT, InTalent-Inditex and Prog.~Atracci\'on Talento CM (Spain);
SRC (Sweden);
the Leverhulme Trust, the Royal Society and UKRI (United Kingdom).

%\input{supplementary}

%\input{appendix}

% This should be taken out in the final paper
%\input{supplementary-app}

\clearpage
\addcontentsline{toc}{section}{References}
%\setboolean{inbibliography}{true}
\bibliographystyle{LHCb}
\bibliography{main,standard,LHCb-PAPER,LHCb-CONF,LHCb-DP,LHCb-TDR}

\newpage
% LHCb collaboration author list
% Data extracted on March 12th, 2026 at 10:31am for paper reference LHCb-PAPER-2025-042
\centerline
{\large\bf LHCb collaboration}
\begin
{flushleft}
\small
R.~Aaij$^{38}$\lhcborcid{0000-0003-0533-1952},
A.S.W.~Abdelmotteleb$^{58}$\lhcborcid{0000-0001-7905-0542},
C.~Abellan~Beteta$^{52}$\lhcborcid{0009-0009-0869-6798},
F.~Abudin\'en$^{58}$\lhcborcid{0000-0002-6737-3528},
T.~Ackernley$^{62}$\lhcborcid{0000-0002-5951-3498},
A. A. ~Adefisoye$^{70}$\lhcborcid{0000-0003-2448-1550},
B.~Adeva$^{48}$\lhcborcid{0000-0001-9756-3712},
M.~Adinolfi$^{56}$\lhcborcid{0000-0002-1326-1264},
P.~Adlarson$^{86}$\lhcborcid{0000-0001-6280-3851},
C.~Agapopoulou$^{14}$\lhcborcid{0000-0002-2368-0147}
% ...

{\footnotesize \it

% ...
$^{14}$Universit{\'e} Paris-Saclay, CNRS/IN2P3, IJCLab, Orsay, France\\
% ...
$^{38}$Nikhef National Institute for Subatomic Physics, Amsterdam, Netherlands\\
% ...
$^{48}$Instituto Galego de F{\'\i}sica de Altas Enerx{\'\i}as (IGFAE), Universidade de Santiago de Compostela, Santiago de Compostela, Spain\\
% ...
$^{52}$Physik-Institut, Universit{\"a}t Z{\"u}rich, Z{\"u}rich, Switzerland\\
% ...
$^{56}$H.H. Wills Physics Laboratory, University of Bristol, Bristol, United Kingdom\\
% ...
$^{58}$Department of Physics, University of Warwick, Coventry, United Kingdom\\
% ...
$^{61}$School of Physics and Astronomy, University of Glasgow, Glasgow, United Kingdom\\
$^{62}$Oliver Lodge Laboratory, University of Liverpool, Liverpool, United Kingdom\\
% ...
$^{70}$Syracuse University, Syracuse, NY, United States\\
% ...
$^{86}$Department of Physics and Astronomy, Uppsala University, Uppsala, Sweden, associated to $^{61}$\\
% ...
}
\end{flushleft}

\end{document}